\documentclass[12pt]{article}

\usepackage{epsf,amsfonts,hyperref}
\usepackage{cite}
\renewcommand{\appendix}[1]{
    \addtocounter{section}{1}
    \setcounter{equation}{0}
    \renewcommand{\thesection}{\Alph{section}}
    \section*{Appendix \thesection\protect\indent #1}
    \addcontentsline{toc}{section}{Appendix \thesection\ \ \ #1}
}
\newcommand\encadremath[1]{\vbox{\hrule\hbox{\vrule\kern8pt
\vbox{\kern8pt \hbox{$\displaystyle #1$}\kern8pt}
\kern8pt\vrule}\hrule}}
\def\enca#1{\vbox{\hrule\hbox{
\vrule\kern8pt\vbox{\kern8pt \hbox{$\displaystyle #1$}
\kern8pt} \kern8pt\vrule}\hrule}}

\newcommand\figureframex[3]{
\begin{figure}[bth]
\hrule\hbox{\vrule\kern8pt
\vbox{\kern8pt \vbox{
\begin{center}
{\mbox{\epsfxsize=#1.truecm\epsfbox{#2}}}
\end{center}
\caption{#3}
}\kern8pt}
\kern8pt\vrule}\hrule
\end{figure}
}
\newcommand\figureframey[3]{
\begin{figure}[bth]
\hrule\hbox{\vrule\kern8pt
\vbox{\kern8pt \vbox{
\begin{center}
{\mbox{\epsfysize=#1.truecm\epsfbox{#2}}}
\end{center}
\caption{#3}
}\kern8pt}
\kern8pt\vrule}\hrule\end{figure}
}

\renewcommand{\thesection}{\arabic{section}}

\makeatletter
\@addtoreset{equation}{section}
\makeatother
\newtheorem{theorem}{Theorem}[section]

\newtheorem{remark}{Remark}[section]
\newtheorem{proposition}{Proposition}[section]
\newtheorem{lemma}{Lemma}[section]
\newtheorem{corollary}{Corollary}[section]
\newtheorem{definition}{Definition}[section]
\newcommand{\eq}[1]{(\ref{#1})}
\def\br{\begin{remark}\rm\small}
\def\er{\end{remark}}
\def\bt{\begin{theorem}}
\def\et{\end{theorem}}
\def\bd{\begin{definition}}
\def\ed{\end{definition}}
\def\bp{\begin{proposition}}
\def\ep{\end{proposition}}
\def\bl{\begin{lemma}}
\def\el{\end{lemma}}
\def\bc{\begin{corollary}}
\def\ec{\end{corollary}}
\def\beaq{\begin{eqnarray}}
\def\eeaq{\end{eqnarray}}
\newcommand{\proof}[1]{{\noindent \bf proof:}\par
{#1} $\square$}

\newcommand{\beq}{\begin{equation}}
\newcommand{\eeq}{\end{equation}}
\newcommand{\bea}{\begin{eqnarray}}
\newcommand{\eea}{\end{eqnarray}}

\renewcommand{\and}{{\qquad {\rm and} \qquad}}

\newcommand{\virg}{{\qquad , \qquad}}

 \newcommand{\Tr}{{\,\rm Tr}\:}
\newcommand{\tr}{{\,\rm tr}\:}

\newcommand{\Res}{\mathop{\,\rm Res\,}}

\newcommand{\td}[1]{{\tilde{#1}}}

\renewcommand{\d}{{{\partial}}}

\newcommand{\Pint}{{\int\kern -1.em -\kern-.25em}}

\newcommand{\ovl}{\overline}

\newcommand{\pbar}{{\overline{p}}}
\newcommand{\qbar}{{\overline{q}}}

\title{{\bf Free energy topological expansion for the 2-matrix model } \vspace{.5cm}}
\author{{\bf L. Chekhov},\thanks{E-mail: \ chekhov@mi.ras.ru}
\date{ } \\ {\small
{\it Steklov Mathematical Institute, ITEP}, and {\it Poncelet Laboratoire, Moscow, Russia}}\\
{\bf B. Eynard},\footnote{E-mail: \ eynard@saclay.cea.fr }
and {\bf N.\ Orantin}\hspace*{0.05cm}\footnote{ E-mail: orantin@cea.fr }\\
\date{ } \\
{\small {\it Service de Physique Th\'{e}orique de Saclay,}}
\\
{\small {\it F-91191 Gif-sur-Yvette Cedex, France.}}
}

\begin{document}

\maketitle

\vspace{-10.5cm}

\begin{center}
\hfill SPhT-T06/016\\
\hfill ITEP/TH-05/06\\
\end{center}

\vspace{7.5cm}

\begin{abstract}
We compute the complete
topological expansion of the formal hermitian two-matrix model. For
this, we refine the previously formulated diagrammatic rules
for computing the $1\over N$ expansion of the nonmixed
correlation functions and give a new formulation  of the spectral
curve. We extend these rules obtaining a closed formula for correlation
functions in all orders of topological expansion. We then integrate it
to obtain the free energy in terms of
residues on the associated Riemann surface.
\end{abstract}

%






\section{Introduction}

Matrix models is a fascinating topic unifying many otherwise seemingly unrelated
disciplines, including integrable systems, conformal theories,
topological expansion, etc~\cite{Mehta,ZJDFG}. However, only
recently the understanding came that geometry itself
plays a crucial role in constructing perturbative solutions to matrix models.
Dijkgraaf and Vafa~\cite{DV} studies in the one-matrix model (1MM), together with subsequent
observations~\cite{KazMar,MarcoF,ChMir,ChMarMirVas} based on the idea of~\cite{Kri} that topological hierarchies
are encoded already in the planar limit of the solutions to the Hermitian one-matrix model
with multiple-connected support of eigenvalues, led to understanding of a crucial role of
the corresponding spectral curve in constructing explicit solutions.

The first attempts of computing the subleading terms in the $1/N^2$ expansion, where based on loop equations \cite{AkAm, ACKM},
and allowed the authors of \cite{ACKM} to find the first few terms of the expansion of the free energy, only in the 1-cut case of the 1-hermitean matrix model.

In the begining of the 2000's, a geometrical approach, supported by using the master loop equation, resulted in constructing solutions in the first subleading
order of the 1MM in the mutlticut case~\cite{eynm2mg1,Chekh, EKK,WZ1}.

Then, a Feynmann-like diagrammatic technique was invented in \cite{eynloop1mat}, which allowed to reformulate the loop equations themselves in a proper geometric way, and it became possible to compute all correlation functions of the 1-MM to all orders in the $1/N^2$ expansion \cite{eynloop1mat}.
However, the diagrammatic technique of \cite{eynloop1mat} could not be applied directly to the expansion of the free energy.
The topological expansion of the free energy to all orders was found with a refined diagrammatic technique in \cite{ChEy}, where the key ingredient was the homogenity property of the free energy to all orders.

In parallel, the hermitian two-matrix model (2MM) solutions have been obtained, at early stages, in the
planar limit of the $1/N$-expansion~\cite{staudacher,eynmultimat}. It was however almost immediately observed
\cite{Bertola,eynm2m,eynmultimat} that the 2MM solution in the planar limit enjoys the same geometrical properties as the
1MM solution; only the spectral curve becomes an arbitrary algebraic curve, not just an hyperellitic curve
arising in the 1MM case. The subsequent progress was however hindered by that the corresponding master
loop equation in the 2MM case cannot immediately be expressed in terms of correlation functions alone. Nevertheless,
using the geometrical properties of this equation, the solution in the first subleading order has been constructed
\cite{eynm2mg1,EKK} on the base of knowledge of Bergmann tau function on Hurwitz spaces. However, the general belief
arose that the 2MM case should not be very much different from the 1MM case, that is, we expect to find
all ingredients of the 1MM solution in the 2MM case. Next step towards constructing
the topological expansion of the 2MM was performed in~\cite{EyOran}, where the first variant of the
diagrammatic technique for the correlation functions in the 2MM case was constructed.
In the present paper we improve the technique of~\cite{EyOran} (actually
effectively simplifying it) to accommodate the action of the loop insertion operator. In fact, we
demonstrate that the diagrammatic technique for the 2MM case is even closer to the one in the 1MM case
that was before: in particular, we need only three-valent vertices, and the additional operator $H$ we need
to obtain the free energy turns out to be of the same origin as the one in the 1MM.

The paper is organized as follows. In Sec.~2, we collect all the definitions, algebraic-geometrical
notation, and facts about loop equations and filling fractions we need in what follows. In Sec.~3, we provide a
new formula for the spectral curve, which results in the new diagrammatic rules formulated in Sec.~4. In the same
section, we express the action of the loop insertion operator in terms of our diagrammatic technique, which
makes the construction closed as regarding the nonmixed multipoint correlation functions. We introduce the ``integration''
operator $H$ in Sec.~5 and, using this operator, present the diagrammatic technique that enables us to construct the
complete topological expansion for the 2MM to all orders except the subleading order.
But for this latter, the answer has been found in~\cite{EKK}, so we eventually formulate a
complete procedure for constructing the topological expansion in the 2MM.
In Appendix A, we extend our technique to calculating the first mixed correlation function; higher mixed correlation
functions need further refining of this technique, which is beyond the scope of this publication. In Appendix~B, we prove
the symmetricity of the free-energy expression w.r.t. interchanging $x$ and $y$ variables, which is crucial for the
proper integration of the expression for the first mixed correlation function.

\section{Definitions and algebraic-geometrical notation}

\subsection{Definition of the model}

We study the formal two-matrix model \cite{Kazakov} and compute the free energy
${\cal F}$ of this model in the asymptotic ${1\over N}$-expansion.
The partition function $Z$ is the formal matrix integral
\beq
\label{defZ}
Z:=\int_{H_N\times H_N} dM_1 dM_2\, e^{-{1\over
\hbar} Tr(V_1(M_1) + V_2(M_2) - M_1 M_2 )} = e^{-{\cal F}},
\eeq
where $M_1$ and $M_2$ are two $N \times N$ Hermitian matrices,
$dM_1$ and  $dM_2$ are the products of Lebesgue measures of the
real components of $M_1$ and $M_2$, $\hbar={T\over N}$ is a formal
expansion parameter and $V_1$ and $V_2$ are two polynomial
potentials of respective degrees $d_1+1$ and $d_2+1$
\beq
\label{defVpot}
V_1(x) = \sum_{k=1}^{d_1+1} {t_k\over k} x^k
\virg V_2(y) = \sum_{k=1}^{d_2+1} {\td{t}_k\over k} y^k .
\eeq
Formal integral means that it is computed order by order in powers
of the $t_k$'s (see section \ref{fillfrac} or \cite{eynhabilit}).
We consider polynomial potentials here only for simplicity of notations,
but it is clear that the whole method presented in this paper extends to "rational potentials" (i.e. such that $V'_1$ and $V'_2$ are rational functions),
and we expect it to extend to the whole semiclassical setting \cite{marcosemipot} including hard edges as well.

One can expand the free energy as well as all the correlation
functions of this model in a $\hbar$ series assuming $\hbar$ to the order of the
reciprocal matrix size; this
procedure is customarily called the {\em topological expansion}
pertaining to the fat-graph representation for formal integrals~(\cite{thooft, BIPZ, ZJDFG}).

The topological expansion of the Feynman diagrams series reads, in
terms of the free energy,
\beq
{\cal F} = {\cal F} (N,T,t_1, t_2,
\dots , t_{d_1+1}, \td{t}_1, \td{t}_2, \dots , \td{t}_{d_2+1}) =
\sum_{h=0}^{\infty} \hbar^{2h-2}
{\cal F}^{(h)}(T,\dots) .
\eeq
We find a
general formula for ${\cal F}^{(h)}$ for any positive integer $h$.
Actually, we address the problem for $h\geq 2$; the solutions for
$h=0$ (\cite{Kri, MarcoF}) and $h=1$ (\cite{EKK}) are already known.

\subsection{Notations}

\subsubsection{Variable sets}

We consider functions of many variables $x_1,x_2,x_3,\dots$, or of
a subset of those variables. For this, we introduce the following
notation:

Let $K$ be a set of $k$ integers:
\beq
K=(i_1,i_2,\dots, i_k).
\eeq
Let $k=|K|$ denote the length (or cardinality) of $K$. For
any $j\leq |K|$, let $K_j$ denote the set of all $j-$upples (i.e.,
subsets of length $j$) contained in $K$:
\beq
K_j:=\{J \subset K
\,\,\, , \,\, |J|=j \} .
\eeq
We define the following $k-$upple of complex numbers:
\beq
{\bf x}_K:=(x_{i_1},x_{i_2},\dots, x_{i_k})
\virg
d{\bf x}_K:= \prod_{j=1}^k dx_j.
\eeq

\subsubsection{Resolvents}

For given integers $k$ and $l$, we define the resolvents:
\bea
\overline{w}_{|K|,|L|}({\bf x}_K,{\bf y}_L)
&:=& {\hbar}^{2-|K|-|L|}\left< \prod_{i=1}^{|K|} \tr{1\over x_{i}-M_1} \prod_{i=1}^{|L|} \tr{1\over y_{i}-M_2}\right>_{conn} \cr
&=& -\hbar^{2} {\partial \over \partial V_2(y_{|L|})} \dots {\partial \over \partial V_2(y_{1})}
{\partial \over \partial V_1(x_{|K|})} {\partial \over \partial V_1(x_{|K|-1})} \dots {\partial {\cal F} \over \partial V_1(x_1)}\cr
&& \qquad \qquad + {\delta_{|K|,1}\delta_{|L|,0}\over x_1}+{\delta_{|K|,0}\delta_{|L|,1}\over y_1} \cr
\eea
with the
formal loop insertion operators
\beq
{\partial \over \partial
V_1(x)} = -\sum_{j=1}^{\infty} {j \over x^{j+1}} {\partial \over
\partial t_j} \;\;\; \hbox{and} \;\;\;
 {\partial \over \partial V_2(x)} = -\sum_{j=1}^{\infty} {j \over
y^{j+1}} {\partial \over
\partial \tilde{t}_j} .
\eeq
We also introduce the polynomials in y:
\beq
\overline{u}_k(x,y;x_{|K|}) := \hbar^{1-k}\left< \tr {1\over
x-M_1}\,{V'_2(y)-V'_2(M_2)\over y-M_2}\,\prod_{r=1}^{|K|}
\tr{1\over x_{r}-M_1}\right>_{conn},
\eeq
where the subscript $conn$ denotes the connected component.

For convenience, we renormalize the two-point functions:
\beq\label{defw2renorm}
w_{|K|,|L|}({\bf x}_K,{\bf y}_L) = \overline{w}_{|K|,|L|}({\bf
x}_K,{\bf y}_L) +{\delta_{|K|,2}\delta_{|L|,0}\over
(x_1-x_2)^2}+{\delta_{|K|,0}\delta_{|L|,2}\over (y_1-y_2)^2}
\eeq
and their polynomial correspondent:
\beq
u_k(x,y;x_K) :=
\overline{u}_k(x,y;x_K) - \delta_{k,0} (V'_2(y)-x) .
\eeq

We consider the $\hbar^2$-expansions of the above quantities:
\beq
\label{defwklh}
w_{|K|,|L|}({\bf x}_K,{\bf y}_L) =
\sum_{h=0}^\infty \hbar^{2h} \,w_{|K|,|L|}^{(h)}({\bf x}_K,{\bf
y}_L) ,
\eeq

\beq\label{defuklh} u_k(x,y;x_K) = \sum_{h=0}^\infty
\hbar^{2h}\,u_k^{(h)}(x,y;x_K) .
\eeq

We also need the functions closely related to the algebraic
structure of the problem:
\beq\label{defY}
Y(x) := V_1'(x) - \overline{w}_1(x)
\eeq
and
\beq
P(x,y) := \hbar\, \left< \tr
{V'_1(x)-V'_1(M_1)\over x-M_1}\,{V'_2(y)-V'_2(M_2)\over
y-M_2}\right>_{conn} .
\eeq

The latter function together with all its terms of
$\hbar^2$-expansion, is a polynomial of degree $d_1-1$ in $x$ and
$d_2-1$ in $y$.

\subsubsection{The master loop equation}
Among many different ways to solve matrix models~\cite{Mehta},
addressing the formal 2-matrix model problem, we choose the
so-called loop equations~\cite{staudacher,Virasoro} encoding the matrix integral in \eq{defZ}
to be invariant under special changes of variables. They
correspond to the generalization of the Virasoro constraints of
the one-matrix model, i.e. the W-algebra.

Considering a particular change of variables, we come to the {\em
master loop equation}~\cite{eynm2mg1}:

\beq\encadremath{\label{mastloop1} E(x,y) = (y-Y(x)) u_0(x,y) +
\hbar^2 u_1(x,y;x) },\eeq where
\beq\label{defE}
E(x,y) = (V_1'(x)-y) (V_2'(y)-x) - P(x,y) + T
\eeq
is a polynomial in $x$ and $y$ defining the {\em spectral curve}.

Considering the 't Hooft expansion in orders of $\hbar^2$, for any
$h \geq 1$, we have: \bea\label{mastloop2} E^{(h)}(x,y) &=&
(y-Y(x)) u_0^{(h)}(x,y) + w_{1,0}^{(h)}(x) u_0^{(0)}(x,y) \cr && +
\sum_{m=1}^{h-1} w_{1,0}^{(m)}(x) u_0^{(h-m)}(x,y) + {\partial
\over
\partial V_1(x)} u_0^{(h-1)}(x,y), \cr \eea
where $E^{(h)}(x,y)$ is the $h$'th term in the $\hbar^2$-expansion
of the spectral curve $E(x,y)$.

\subsubsection{Algebraic geometry notations}
In this section, we recall the algebraic-geometrical pattern of
our problem (see more details in \cite{EyOran,Fay,Farkas}).

To leading order, the {\em master loop equation} \eq{mastloop2}
reduces to an algebraic equation \beq\label{master}
E^{(0)}(x,Y(x))=0 \eeq with $E^{(0)}(x,y)$ being a polynomial of
degree $d_1+1$ in $x$ and $d_2+1$ in $y$.

We parameterize the algebraic curve $E^{(0)}(x,y)=0$ implied by
\eq{master} by a running point $p$ on the corresponding compact
Riemann surface ${\cal E}$. We therefore define two analytical
meromorphic functions $x(p)$ and $y(p)$ on ${\cal E}$ such that:
\beq E^{(0)}(x,y)=0 \Leftrightarrow \exists ! \, p \in {\cal E}
\,\,\,\,\, x=x(p) \,\, , \,\, y=y(p) .\eeq The functions $x$ and
$y$ are not bijective. Indeed, since $E^{(0)}(x,y)$ has a degree
$d_2+1$ in $y$, it admits $d_2+1$ solutions for a given $x$; that
is we have $d_2+1$ points $p$ on $ {\cal E} $ such that
$x(p)=x$. Thus, the Riemann surface is made of $d_2+1$ $x$-sheets,
or respectively, of $d_1+1$ $y$-sheets. We then denote \beq x(p) =
x \Leftrightarrow p = p^{(j)}(x) \,\,\,\,\, \hbox{for} \,\,\,\,
j=0,\dots,d_2, \eeq

\beq y(p) = y \Leftrightarrow p = \td{p}^{(j)}(x) \,\,\,\,\,
\hbox{for} \,\,\,\, j=0,\dots,d_1. \eeq

Among the different $x$-sheets (resp. $y$-sheet), there exists
only one where $y(p) \sim_{x(p) \to \infty} V_1'(x(p)) -{T \over
x(p)} + O(x^2(p))$ (resp. $x(p) \sim_{y(p) \to \infty} V_2'(y(p))
-{T \over y(p)} + O(y^2(p))$). We call it the physical sheet, and
it bears the superscript 0.

\bigskip
{\bf Genus and cycles.} The curve ${\cal E}$ is a compact
Riemann surface of finite genus $g\leq d_1 d_2 -1$\footnote{This
genus g must not be confused with the genus indicating the order
of the topological expansion} . We do not require $g$ to be equal
to $d_1d_2-1$, that is, we allow double points on the
corresponding Riemann surface. We choose $2g$ canonical cycles as
${\cal A}_i$, ${\cal B}_i$, $i=1,\dots, g$, such that: \beq {\cal
A}_i\cap {\cal A}_j=0 \virg {\cal B}_i\cap {\cal B}_j=0 \virg
{\cal A}_i\cap {\cal B}_j=\delta_{ij} .\eeq

\bigskip
{\bf Branch points.} The $x$-branch points $\mu_{\alpha}$, $\alpha
=1,\dots, d_2+1+2g$, are the zeroes of the differential $dx$,
respectively, the $y$-branch points $\nu_{\beta}$, $\beta
=1,\dots, d_1+1+2g$, are the zeroes of $dy$. We assume here that
all branch points are simple and distinct. Note also that
$E^{(0)}_y(x(p),y(p))$ vanishes (simple zeroes) at the branch
points (it vanishes in other points as well).

\smallskip
A branch point is a point where two sheets of the Riemann surface meet.
If $p$ is a point on the Riemann surface near a branch point, there is another point $p^i$, which we note $\pbar$, near the same branch point.
In other words:
\beq \forall \alpha \;\;\; \exists ! \overline{p} \neq p \;\;\;
\hbox{such that}\;\;\; x(\overline{p}) = x(p) \;\;\; \hbox{and}
\;\;\; y(\overline{p}) \to_{p \to \mu_\alpha} y(p) \eeq

Let us emphasize that the point $\pbar$ and the corresponding sheet, depend on which branch point we are considering.

\bigskip
{\bf Bergmann kernel.} On the Riemann surface ${\cal{E}}$, we have
a unique Abelian bilinear differential $B(p,q)$, with one double
pole at $p=q$ such that \beq\label{defB} B(p,q)\mathop\sim_{p\to
q} {dx(p)dx(q)\over (x(p)-x(q))^2}+{\rm finite} \quad {\rm and}
\quad \forall i \,\,\,\oint_{{p\in{\cal A}_i}} B(p,q) = 0 .\eeq It
is symmetric, \beq B(p,q)=B(q,p), \eeq it can be expressed in
terms of theta-functions \cite{Fay, Farkas}, and depends only on
the complex structure of ${\cal E}$.

\bigskip
{\bf Abelian differential of the third kind.}

On the Riemann surface ${\cal E}$, there exists a unique Abelian
differential of the third kind $dS_{q,r}(p)$, with two simple
poles at $p=q$ and at $p=r$, such that \beq\label{defdS} \Res_{p
\to q} dS_{q,r}(p) = 1 = -\Res_{p \to r} dS_{q,r}(p) \quad {\rm
and} \quad \forall i \,\,\,\oint_{{{\cal A}_i}} dS_{q,r}(p) = 0.
\eeq

Notice that the Abelian differential of the third kind and the
Bergmann kernel are linked by
\beq
dS_{q,r}(p) = \int_{\xi=r}^q B(\xi,p)
\;\;\; \hbox{and} \;\;\;
B(p,q) = d_q  \left( dS_{q,r}(p) \right).
\eeq
where the contour of integration is a line from $r$ to $q$, which does not cross any ${\cal A}$ or ${\cal B}$ cycle.

\smallskip
Given a branch point $\mu_\alpha$, and a point $q$ in the vicinity of $\mu_\alpha$, we introduce the following notation:
\beq
dE_{q,\qbar}(p) = \int_{\qbar}^q B(\xi,p),
\eeq
where now, the integration path is chosen as the shortest path between $q$ and $\qbar$, i.e. a path which lies in a small vicinity of  $\mu_\alpha$.
That definition differs from the one above.
If the branchpoint $\mu_\alpha$ is surrounded by contour ${\cal A}_i$, we have:
\beq
dE_{q,\qbar}(p) = dS_{q,\qbar}(p) + \oint_{{\cal B}_i} B(p,\xi)
\eeq
The main property of that $dE_{q,\qbar}(p)$, is  that it vanishes at $q=\qbar$, i.e. it vanishes at the branch point $\mu_\alpha$.

\bigskip
{\bf Correlation functions on the Riemann surface}

Given the algebraic curve,
we see that we can redefine the correlation functions \eq{defwklh} and \eq{defuklh} more precisely.
Indeed, they are defined only as formal series as their arguments ${\bf x}_K$ and ${\bf y}_L$ tend to infinity.
The loop equations show that these formal series are in fact algebraic functions and hence have a finite radius of convergency,
with cuts beyond the radius. As functions of the $x$ and $y$ variables, they are multivalued.

If instead of writing them as functions of $x$ or $y$, we write them as functions on the Riemann surface, they become monovalued.
This is the reason why we prefer to introduce the following notation for the correlation functions, as meromophic differential forms on the Riemann surface:
\beq W_{|K|,|L|}({\bf p}_K,{\bf q}_L)
:= w_{|K|,|L|}(x({\bf p}_K),y({\bf q}_L)) dx({\bf p}_K) dy({\bf
q}_L),\eeq
and
\beq U_{|K|}(p,y;{\bf p}_K) := u_{|K|}(x(p),y;x({\bf p}_K))
dx({\bf p}_K), \eeq where $p_i$'s and $q_j$'s are points on the
surface ${\cal E}$ whose images $x(p_i)$ or $y(q_i)$ are complex numbers. $y$ is a complex number.

\br
The interpretation of those correlation functions as generating series for the moments $<\prod_i \Tr M_1^{k_i} \prod_j \Tr M_2^{l_j}>_{\rm c}$,
corresponds to the situation where all $p_i$'s are in the $x$-physical sheet, in a vicinity of $\infty_x$, and all $q_j$'s are in the $y$-physical sheet in a vicinity of $\infty_y$.
\er

\bigskip
{\bf 2 point function}

It is well known ( see for instance \cite{Ak96, Kri, Kos, Bertola, KazMar, eynm2mg1, eynmultimat}) that, to leading order, the 2-point function $W_{2,0}$ is the Bergmann kernel:
\beq\label{W20B}
W^{(0)}_{2,0}(p,q) = B(p,q)
\eeq

The non-renormalized 2 point function:
\beq
\ovl{W}_{2,0}(p,q)=W_{2,0}(p,q)-{dx(p)dx(q)\over (x(p)-x(q))^2}
\eeq
is finite at $p=q$.

\subsection{Loop equations and fixed filling fractions}
\label{fillfrac}

We have showed that to leading order, the 1-point function $Y^{(0)}(x)$ obeys an algebraic equation \eq{master}:
\beq E^{(0)}(x,Y(x))=0, \eeq
where
\beq
E^{(0)}(x,y) = (V_1'(x)-y) (V_2'(y)-x) - P^{(0)}(x,y) + T
\eeq
But so far we have not discussed how to determine the polynomial $P^{(0)}(x,y)$.

As this was extensively discussed in the literature, we only briefly summarize it below.

\smallskip
We need $d_1 d_2-1$ equations to fix the $d_1 d_2-1$ unknown coefficients of $P^{(0)}$.
Those additional $d_1 d_2-1$ equations do not come from any loop equation, so we need an independent hypothesis related to the precise definition of our matrix model.
In a sense, loop equations express the invariance of the integral under reparameterizations, independently on the integration paths.
Additional equations are those that depend on the choice of integration path.

For arbitrary integration paths, the $\hbar$ expansion may not exist.
The choice of integration method we use in this paper, corresponds to the so-called {\bf formal matrix model, with fixed filling fractions}.

\medskip
We consider here the problem of a formal matrix model, i.e. the
one represented by the formal power series expansion of a matrix
integral, where the non-quadratic terms in the potentials $V_1$
and $V_2$ are treated as perturbations near quadratic potentials.
Such a perturbative expansion can be performed only near local
extrema of $V_1(x)+V_2(y)-xy$, i.e. near the points $(\xi_i,\eta_i)$, $i=1,\dots,d_1d_2$, such that
\beq
V'_1(\xi_i)=\eta_i \;\;\; \hbox{and} \;\;\; V'_2(\eta_i)=\xi_i
\eeq which has in general $d_1 d_2$ solutions.
Therefore, perturbative expansion can be performed near matrices of the form:
\beq
\ovl{M_1} = {\rm diag}\,(\mathop{\overbrace{\xi_1,\dots,\xi_1}}^{n_1},\mathop{\overbrace{\xi_2,\dots,\xi_2}}^{n_2},\dots,\mathop{\overbrace{\xi_{d_1 d_2},\dots,\xi_{d_1 d_2}}}^{n_{d_1 d_2}}) ,
\eeq
\beq
\ovl{M_2} = {\rm diag}\,(\mathop{\overbrace{\eta_1,\dots,\eta_1}}^{n_1},\mathop{\overbrace{\eta_2,\dots,\eta_2}}^{n_2},\dots,\mathop{\overbrace{\eta_{d_1 d_2},\dots,\eta_{d_1 d_2}}}^{n_{d_1 d_2}}),
\eeq
such that $\sum_{i=1}^{d_1 d_2} n_i=N$.
The perturbative  integral is computed by writing $M_1=\ovl{M_1}+\delta M_1$ and  $M_2=\ovl{M_2}+\delta M_2$,
by expanding higher order terms (cubic and higher) in $\delta M_1$ and $\delta M_2$,
and by treating quadratic terms in $\delta M_1$ and $\delta M_2$ as  gaussian integrals.

In other words, a formal integral can be computed as soon as we have chosen a vacuum around which to make a perturbative expansion.
The choice of a vacuum is equivalent to choosing a partition of $N$ into $d_1 d_2$ parts:
\beq
\sum_{i=1}^{d_1 d_2} n_i=N .
\eeq

It is easy to see that if we truncate the perturbative expansion to any finite order, we have:
\beq
{1\over 2i\pi}\oint_{{\cal C}_i} \left<\tr {1\over x-M_1}\right> dx = {1\over 2i\pi}\oint_{{\cal C}_i} \tr {1\over x-\ovl{M_1}} dx = -n_i,
\eeq
where ${\cal C}_i$ is  a small direct circle around the point $\xi_i$ in the complex plane.
And thus:
\beq
{1\over 2i\pi}\oint_{{\cal C}_i} w_{1,0}(x) dx = - n_i \hbar:=-\epsilon_i .
\eeq
If we do not truncate to a finite order, all functions become algebraic, with cuts, and the contours ${\cal C}_i$ are then enhanced to finite contours around the cuts, and up to
a redefinition of the contours,
the {\em filling fractions} are the ${\cal A}$-cycle integrals. We define:
\beq\label{deffillingfractions}
{1\over 2i\pi}\oint_{{\cal A}_i} y dx =-{1\over 2i\pi}\oint_{{\cal A}_i} x
dy =\epsilon_i
\eeq
We call $\epsilon_i$'s the filling fractions, and they are given new parameters (moduli) of the model.
In what follows, we consider them to be independent of the potential or on any other parameter.
Let us notice that the such defined $\epsilon_i$'s are not necessarily of the form $n_i/N$ with $n_i$ a positive integer, but they can be arbitrary complex numbers, provided that:
\beq
\sum_i \epsilon_i=t=N\hbar
\eeq

In particular, since all correlation functions
$w_{k,0}(x_1,\dots,x_k)$ are obtained by derivation of $w_{1,0}$ with
respect to the potential $V_1$, we have:
\beq\label{vanishingAcycles} {1\over 2i\pi}\oint_{{\cal A}_i}
w_{k,0}(x_1,\dots,x_k) dx_1 =0 .\eeq

Eq.(\ref{deffillingfractions}) together with the large $x$ and $y$
behaviors, suffices for determining completely all the
coefficients of the polynomial $E^{(0)}(x,y)$, and thus the
leading large $N$ resolvents $w_{1,0}(x)$ and $w_{0,1}(y)$.

\br
It is easy to see that if we truncate the perturbative gaussian integral to any finite order, the result is a polynomial in $1/N^2$, and thus, formal matrix models always have a $1/N^2$ expansion.
This comes from  the fact that, to any finite order, the perturbative gaussian integral is the generating function for counting discrete surfaces made of a finite bounded number of polygons,
thus having a maximal Euler characteristic (power of $N$).
\er

\br
We emphasize again the formality of our model: in principle, the ${\cal A}$-cycles do not
necessarily lie in the physical sheet, so we would need additional assumptions to segregate physically meaningful models.
Recall that the $\epsilon_i$'s  can be arbitrary complex numbers.
\er

\br
We have $d_1 d_2$ non vanishing contour integrals, but they are not all independent. Their sum can be deformed as a contour around $\infty$.
Thus, we only have $d_1 d_2-1$ independent non vanishing contour integrals, and thus the maximal genus of our curve is $d_1 d_2-1$.
The number of independent filling fractions is $d_1 d_2-1$, due to the constraint on their sum.
\er

\bigskip
{\bf Double points}
The Riemann surface ${\cal E}$ may have double points.

A point $p\in{\cal E}$ is a double point iff:
\beq
\exists q\neq p,\,\,\, x(p)=x(q)\,\, \hbox{and}\,\,\, y(p)=y(q).
\eeq
Double points are such that both derivatives $E_x$  and $E_y$ vanish, and thus the curve has two tangents at such points.
The differential forms $dx$ and $dy$ do not vanish at double points.

We need an extra asumption to deal with double points.

We may consider that in the framework of the perturbative gaussian matrix integral seen above, double points correspond to vanishing filling fractions $\epsilon_i=0$.
Thus:
\beq\label{vanishingdbptres}
\oint_{\cal C} w_{1,0}(x)dx = 0
\eeq
where ${\cal C}$ is a contour which encircles the double point.
In other words, we make the assumption that the resolvent has  no residue at double points, to all orders in the $\hbar$ expansion.
We will see below that this vanishing residue condition is sufficient to ensure that correlation functions never have poles at double points.

Another reason to make that assumption is based on the Krichever formula~\cite{Kri}, which involves summation only over branch-points and not over double points.

\section{A new formula for the spectral curve}

The 't Hooft expansion  of the master loop equation \eq{mastloop2}, links correlation functions of genus $h$
to correlation functions of lower genus.
The knowledge of the LHS $E(x,y)$ would then give a way of deriving the whole topological expansion of the $W_{|K|,0}$.

From the definition of the surface ${\cal E}$, one derives the
leading term as:

\bea E^{(0)}(x(p),y(q)) &=& - t_{d_1+1} \times \prod_{i=0}^{d_1}
(x(p)-x(\td{q}^{(i)}(y))) \cr &=& - \td{t}_{d_2+1} \times
\prod_{i=0}^{d_2} (y(q)-y(p^{(i)}(x))) \eea

Actually, not only the leading term, but the whole function admits such a product structure:

\bt
The functions $E(x,y)$ and $U_0(p,y)$ can be written:

\beq\label{expE}
\encadremath{
E(x(p),y) = -\td{t}_{d_2+1}\,"\left<\prod_{i=0}^{d_2} (y-V'_1(x(p))+\hbar \Tr{1\over x(p^i)-M_1}) \right>"
}\eeq
and
\beq\label{expU}
U_{0}(p,y) = -\td{t}_{d_2+1}\,"\left<\prod_{i=1}^{d_2} (y-V'_1(x(p))+ \hbar \Tr{1\over x(p^i)-M_1})  \right>" .
\eeq
where the quotes $"\left<.\right>"$ mean that when we expand into cumulants, each time we find a 2-point function $\ovl{w}_{2,0}$, we replace it by $w_{2,0}=\ovl{w}_{2,0}+{1\over (x_1-x_2)^2}$ as in \eq{defw2renorm}.
\et

In other words, $E(x,y)$ is the $(d_2+1)-$point correlation function, with all points corresponding to the same $x$ in all the $d_2+1$ sheets,
and $U_{0}(p,y)$ is the $d_2$-point correlation function, with all points corresponding to the same $x(p)$ in all sheets but the same as $p$.

\proof{ We prove this theorem by showing that the two sides of the
equalities \eq{expE} and \eq{expU} are defined by the same
recursion relations.

We let $\widetilde{E}(x(p),y)$ and $\td{U}_0(p,y)$ be defined by the respective RHS of \eq{expE} and \eq{expU}.
Consider their topological expansions:
\beq \widetilde{E}(x,y) = \sum_{h=0}^{\infty}
\hbar^{2h} \widetilde{E}^{(h)}(x,y) \;\;\; \hbox{and} \;\;\;
\td{U}_0(p,y) = \sum_{h=0}^{\infty} \hbar^{2h}
\widetilde{U}_0^{(h)}(x(p),y) \eeq

By expanding in cumulants (connected parts), one can observe that:

\bea \widetilde{E}^{(h)}(x(p),y) &=& (y-y(p))
\widetilde{U}_0^{(h)}(p,y) + W_{1,0}^{(h)}(p)
\widetilde{U}_0^{(0)}(p,y) \cr && + \sum_{m=1}^{h-1} W_{1,0}^{(m)}(p)
\widetilde{U}_0^{(h-m)}(p,y) + {\partial \over \partial V_1(x(p))}
\widetilde{U}_0^{(h-1)}(p,y), \cr \eea which has exactly the same
form as \eq{mastloop2}.

Obviously we have $E^{(0)}(x,y)=\widetilde{E}^{(0)}(x,y)$ and $U^{(0)}(p,y)=\widetilde{U}^{(0)}(p,y)$.

Let us now show that, given $E^{(0)}(x,y)$, \eq{mastloop2} admits unique solution.
For this, assume that we know
$E^{(h-1)}(x,y)$ and $U_0^{(m)}(p,y)$ for all $m \leq h-1$ and
prove that \eq{mastloop2} allows computing $E^{(h)}(x,y)$ and
$U_0^{(h)}(p,y)$.

Consider \eq{mastloop2} for $p=q^i$ and $y=y(p)$ for any $i= 0
\dots d_2$. It reads \beq E^{(h)}(x(p),y(p^i)) = \sum_{m=1}^h
W_{1,0}^{(m)}(p^i) U_0^{(h-m)}(p^i,y(p^i)) + {\partial \over
\partial V_1(p^i)} U_0^{(h-1)}(p^i,y(p^i);p^i) .\eeq

By the recursion hypothesis, one knows the RHS totally, so one
knows the LHS. This gives the value of $E^{(h)}(x,y)$ for $d_2+1$
values of $y$. Because $E^{(h)}(x,y)$ is a polynomial in $y$ of
degree $d_2$, an interpolation formula gives its value for any
$y$.
It remains to compute $U_0^{(h)}(p,y)$ which is straightforward due to \eq{mastloop2}.

Thus, because $E^{(h)}(x,y)$ and $\widetilde{E}^{(h)}(x,y)$ obey the same equations with the same initial condition,
we conclude that $E^{(h)}(x,y) = \widetilde{E}^{(h)}(x,y)$ for any $h$.

}

\section{Diagrammatic rules for the correlation functions: new insight.}

Using loop equations, two of the authors have derived two sets of
diagrammatic rules allowing to compute any $W_{|K|,0}$ as residues
on the Riemann surface \cite{EyOran}. One of these theories involved only
cubic vertices but presented the disadvantage of using explicitly
the auxiliary functions $U_{|K|}$, whereas the second one involved
only the $W_{|K|,0}$'s but expressed it through vertices of
valence up to $d_2$.

In this section, we present new diagrammatic rules composed of
cubic vertices and involving only $W$'s whose arguments are in the
vicinity of the branch points. They are much simpler than all
the preceding ones.

\medskip
From \eq{expE} and the definition \eq{defE} of $E(x,y)$, we obtain
the following equation: \beq
\begin{array}{rcl}
-\td{t}_{d_2+1}\,"\left<\prod_{i=0}^{d_2} (y-V'_1(x(p))+\hbar
\Tr{1\over x(p^i)-M_1}) \right>" &=& (V_2'(y)-x(p))(V_1'(x(p))-y)
\cr && - P(x(p),y) +1. \cr \end{array} \eeq

Recall that $P(x,y)$ is a polynomial both in $y$ (of degree $d_2-1$)
{\bf and} in $x$.

The large $y$ expansion of the LHS reads
\bea && - \td{t}_{d_2+1}\,
y^{d_2+1} - \td{t}_{d_2+1}\, \sum_{i=0}^{d_2} \left< \hbar
\Tr{1\over x(p^i)-M_1} - V_1'(x(p)) \right> y^{d_2}\cr && -
\td{t}_{d_2+1}\, \sum_{i=0}^{d_2} \sum_{j\neq i} "\left< \left(
\hbar \Tr{1\over x(p^i)-M_1} - V_1'(x(p)) \right) \times \right.
\cr && \;\;\;\; \left. \times \left( \hbar \Tr{1\over x(p^j)-M_1}
- V_1'(x(p)) \right) \right>"  y^{d_2-1} \cr &&  +O(y^{d_2-2}), \cr
\eea
and the large $y$ expansion of the RHS yields:
\beq
 - \td{t}_{d_2+1}\, y^{d_2+1}  - y^{d_2} ( \td{t}_{d_2} - \td{t}_{d_2+1} V_1'(x(p)) ) - \td{t}_{d_2+1} Q(x(p)) y^{d_2-1} +
 O(y^{d_2-2}),
\eeq
where $Q(x(p))$ is a polynomial in $x$ of degree $d_1-1$.

\smallskip
$\bullet$ Equating the coefficient of $y^{d_2}$ gives
\beq\encadremath{\label{yexp1}
\sum_{i=0}^{d_2} Y(p^i) = V_1'(x(p)) - {\td{t}_{d_2} \over \td{t}_{d_2+1}} },
\eeq
which is a polynomial in $x(p)$.
Expanded to order $h$, this means that
\beq\label{sumWhi}
\sum_{i=0}^{d_2} W^{(h)}_{1,0}(p^i) = 0 \quad \hbox{for}\, h\geq 1.
\eeq
Applying the loop insertion operator to \eq{yexp1} we get:
\beq
\sum_{i=0}^{d_2} W^{(h)}_{2,0}(p^i,q) = \delta_{h,0}\,{dx(p)dx(q)\over (x(p)-x(q))^2},
\eeq
which can also be written:
\beq\label{sumW2ineqj}
\ovl{W}_{2,0}(p^i,q) + \sum_{j\neq i} W_{2,0}(p^j,q) = 0.
\eeq

\smallskip
$\bullet$ Equating the coefficient of $y^{d_2-1}$ gives
\beq\label{yexp2}
{1\over 2}\sum_{i\neq j=0}^{d_2}  Y(p^i)\,Y(p^j)+ \hbar^2 w_{2,0}(p^i,p^j)
= V'_1(x(p)) {\td{t}_{d_2}\over \td{t}_{d_2+1}} + {\td{t}_{d_2-1}\over \td{t}_{d_2+1}} + Q(x(p))
\eeq
where
\beq
Q(x) = \hbar\left<\Tr{V'_1(x)-V'_1(M_1)\over x-M_1}\right>
\eeq
is a polynomial in $x$ of degree $d_1-1$.
Using \eq{yexp1} and \eq{sumW2ineqj} we get:
\beq\label{yexp3}
\sum_{i=0}^{d_2}  Y(p^i)^2 + \hbar^2\ovl{w}_{2,0}(p^i,p^i)
= V'_1(x(p))^2 -2{\td{t}_{d_2-1}\over \td{t}_{d_2+1}}  + \left({\td{t}_{d_2}\over \td{t}_{d_2+1}}\right)^2 -2 Q(x(p))
\eeq
Expanding this equation in $\hbar$, we obtain for $h\geq 1$
\beq\label{yexp4}
\encadremath{
\begin{array}{l}
 2 \sum_{i=0}^{d_2} y(p^i) W_{1,0}^{(h)}(p^i)dx(p) \cr
= {\sum_{i=0}^{d_2} \sum_{m=1}^{h-1} W_{1,0}^{(m)}(p^i) W_{1,0}^{(h-m)}(p^i) } + {\sum_{i=0}^{d_2} \ovl{W}_{2,0}^{(h-1)}(p^i,p^i)}  + 2 Q^{(h)}(x(p)) dx(p)^2 .
\end{array}
} \eeq

We consider this equation when $p$ is near a branch point $\mu_\alpha$, multiply it by ${1 \over 2} {dE_{p,\overline{p}}(q)
\over y(p)-y(\overline{p})} $, take the residues when $p \to \mu_\alpha$ and sum over all the branch points.
Notice that ${1 \over 2} {dE_{p,\overline{p}}(q) \over y(p)-y(\overline{p})}$ is finite at $p=\pbar$, i.e. it has no poles at branch points.

Let us first compute the LHS:
\bea
&& \sum_{\alpha} \Res_{p \to \mu_\alpha}  {dE_{p,\overline{p}}(q)
\sum_{i=0}^{d_2} y(p^i) W_{1,0}^{(h)}(p^i) \over y(p)-y(\overline{p})} \cr
&=& \sum_{\alpha} \Res_{p \to \mu_\alpha}
{dE_{p,\overline{p}}(q) \left[y(p) W_{1,0}^{(h)}(p) + y(\pbar) W_{1,0}^{(h)}(\pbar) + \sum_{p^i\neq p,\pbar} y(p^i) W_{1,0}^{(h)}(p^i) \right] \over y(p)-y(\overline{p})}\cr
&=& \sum_{\alpha} \Res_{p \to \mu_\alpha}
{dE_{p,\overline{p}}(q) \left[y(p) W_{1,0}^{(h)}(p) + y(\pbar) W_{1,0}^{(h)}(\pbar)  \right] \over y(p)-y(\overline{p})}\cr
&=& \sum_{\alpha} \Res_{p \to \mu_\alpha}
{dE_{p,\overline{p}}(q) \left[y(p) W_{1,0}^{(h)}(p) - y(\pbar) (W_{1,0}^{(h)}(p)+ \sum_{p^i\neq p,\pbar} W_{1,0}^{(h)}(p^i)) \right] \over y(p)-y(\overline{p})}\cr
&=& \sum_{\alpha} \Res_{p \to \mu_\alpha}
{dE_{p,\overline{p}}(q) \left[y(p) W_{1,0}^{(h)}(p) - y(\pbar) W_{1,0}^{(h)}(p) \right] \over y(p)-y(\overline{p})}\cr
&=& \sum_{\alpha} \Res_{p \to \mu_\alpha} dE_{p,\overline{p}}(q) \, W_{1,0}^{(h)}(p)  \cr
&=& \sum_{\alpha} \Res_{p \to \mu_\alpha} dS_{p,o}(q) \, W_{1,0}^{(h)}(p) - dS_{\pbar,o}(q) \, W_{1,0}^{(h)}(p)  \cr
&=& \sum_{\alpha} \Res_{p \to \mu_\alpha} dS_{p,o}(q) \, (W_{1,0}^{(h)}(p)-W_{1,0}^{(h)}(\pbar))  \cr
&=& \sum_{\alpha} \Res_{p \to \mu_\alpha} dS_{p,o}(q) \, (2W_{1,0}^{(h)}(p)+ \sum_{p^i\neq p,\pbar} W_{1,0}^{(h)}(p^i))  \cr
&=& 2 \sum_{\alpha} \Res_{p \to \mu_\alpha} dS_{p,o}(q) \, W_{1,0}^{(h)}(p)  \cr
&=& - 2 \Res_{p \to q} dS_{p,o}(q) \, W_{1,0}^{(h)}(p)  \cr
&& \qquad + {1\over 2i\pi}\,\sum_i \left(\oint_{{\cal A}_i} W_{1,0}^{(h)}(p)  \oint_{{\cal B}_i} B(p,q)  + \oint_{{\cal B}_i} W_{1,0}^{(h)}(p)  \oint_{{\cal A}_i} B(p,q) \right) \cr
&=& - 2 \Res_{p \to q} dS_{p,o}(q) \, W_{1,0}^{(h)}(p)   \cr
&=&  2 W_{1,0}^{(h)}(q)   \cr
\eea

Let us compute the RHS in a similar way. It reads
\bea
&& \sum_{\alpha} \Res_{p \to \mu_\alpha}
{{1\over 2}dE_{p,\overline{p}}(q) \over (y(p)-y(\overline{p})) dx(p)} \Big[ \sum_{i=0}^{d_2} \Big( \ovl{W}_{2,0}^{(h-1)}(p^i,p^i)  \cr
&& \qquad + \sum_{m=1}^{h-1} W_{1,0}^{(m)}(p^i) W_{1,0}^{(h-m)}(p^i)  \Big)   +  2 Q^{(h)}(x(p)) dx(p)^2 {\Big]}  \cr
&=& \sum_{\alpha} \Res_{p \to \mu_\alpha}
{{1\over 2}dE_{p,\overline{p}}(q) \over (y(p)-y(\overline{p})) dx(p)} \Big[ \ovl{W}_{2,0}^{(h-1)}(p,p) + \sum_{m=1}^{h-1} W_{1,0}^{(m)}(p) W_{1,0}^{(h-m)}(p)   \cr
&& + \ovl{W}_{2,0}^{(h-1)}(\pbar,\pbar) + \sum_{m=1}^{h-1} W_{1,0}^{(m)}(\pbar) W_{1,0}^{(h-m)}(\pbar) \Big] \cr
&=& 2 \sum_{\alpha} \Res_{p \to \mu_\alpha}
{{1\over 2}dE_{p,\overline{p}}(q) \over (y(p)-y(\overline{p})) dx(p)} \Big[ \ovl{W}_{2,0}^{(h-1)}(p,p) + \sum_{m=1}^{h-1} W_{1,0}^{(m)}(p) W_{1,0}^{(h-m)}(p) \Big]  \cr
\eea

Thus we have, for $h\geq 1$:
\beq W_{1,0}^{(h)}(q) =\sum_{\alpha} \Res_{p \to \mu_\alpha} { {1 \over 2} dE_{p,\overline{p}}(q)  ( \ovl{W}_{2,0}^{(h-1)}(p,p) + \sum_{m=1}^{h-1} W_{1,0}^{(m)}(p) W_{1,0}^{(h-m)}(p) ) \over(y(p)-y(\overline{p})) dx(p)}
\eeq

Using properties \eq{sumWhi} and \eq{sumW2ineqj}, this last equation can be changed into (we add terms which don't have poles at branch points, and thus whose resiudes vanish):
\beq\label{recW1}
W_{1,0}^{(h)}(q) = - \sum_{\alpha} \Res_{p \to \mu_\alpha} { {1 \over 2} dE_{p,\overline{p}}(q)  (W_{2,0}^{(h-1)}(p,\pbar) + \sum_{m=1}^{h-1} W_{1,0}^{(m)}(p) W_{1,0}^{(h-m)}(\pbar) ) \over(y(p)-y(\overline{p})) dx(p)}
\eeq

This is the case $k=0$ of the following theorem (proved below by recursively applying the loop insertion operator $\d/\d V_1$):
\bigskip

{\bf Theorem:} \beq\label{conjecturerecW} \encadremath{
\begin{array}{rcl}
 W_{k+1,0}^{(h)}(q,p_K)
&=&   - \sum_{\alpha} \Res_{p \to \mu_\alpha} {{1\over
2}dE_{p,\pbar}(q)\over (y(p)-y(\pbar))\,dx(p)}\left(
W_{k+1,0}^{(h-1)}(p,\overline{p},p_K) + \right. \cr && \;\;\; +
\left. \sum_{j,m} W_{j+1,0}^{(m)}(p,p_J) \,
W_{k+1-j,0}^{(h-m)}(\overline{p},p_{K-J}) \right) , \cr
\end{array}}\eeq
which can be diagrammatically represented by

\beq
\begin{array}{l}
{\epsfxsize 10cm\epsffile{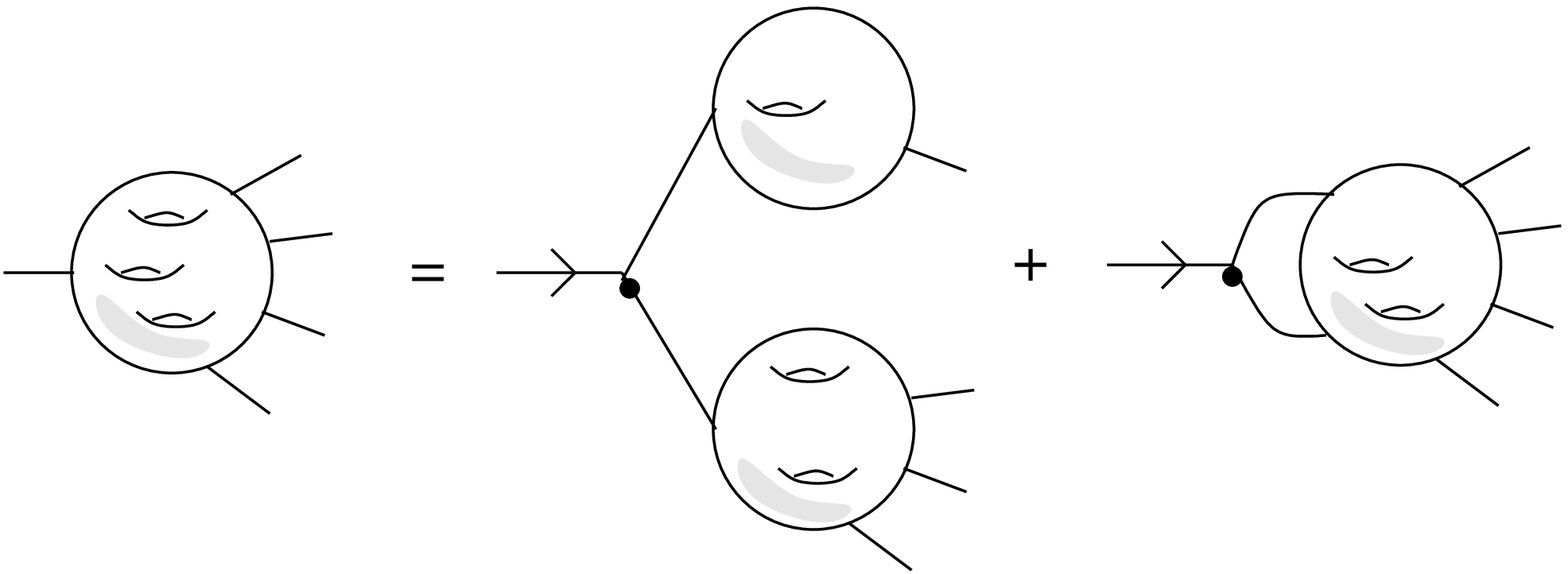}}
\end{array}
\eeq
where a sphere with $h$ holes and $k$ legs is the $k$-point function to order $h$ $W^{(h)}_k$,
and the arrow means the following integration:
\beq\label{vertexrule}
\begin{array}{l}
{\epsfxsize 2.4cm\epsffile{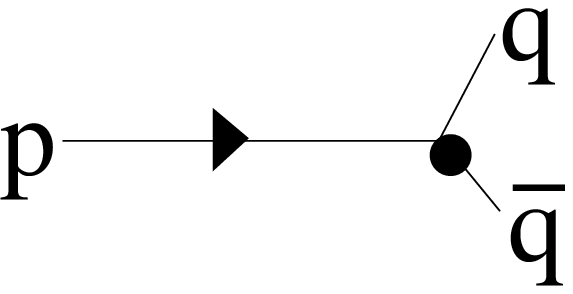}}
\end{array}
=\, \sum_\alpha \Res_{p\to \mu_\alpha} {-{1\over
2}dE_{q,\qbar}(p)\over (y(q)-y(\qbar)) \,dx(q)}
\eeq
This equation is a recursion relation, whose initial condition is given by \eq{W20B}, i.e.
\beq
\begin{array}{l}
{\epsfxsize 2.4cm\epsffile{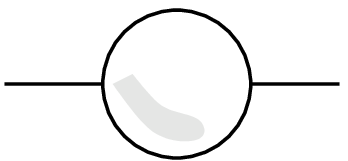}}
\end{array}
 =
\begin{array}{l}
{\epsfxsize 2.4cm\epsffile{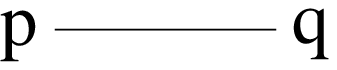}}
\end{array}
=\, B(p,q).
\eeq

\proof{ We have already proved theorem \eq{conjecturerecW} for $k=0$.
Let us now show that the rule \eq{vertexrule} is compatible with the derivation wrt the potential $V_1$.

One preliminary needed formula is the action of the loop insertion
operator on the function $y(p)$. It is well known (see for exemple
\cite{eynm2mg1}) that: \beq {\partial y(p) \over \d V_1(x(r))}\,dx(r) = - {B(p,r)\over dx(p)}
\eeq

First, the action of the loop insertion operator on the Bergmann
kernel gives \cite{EyOran}:
\bea\label{difB}
{\d B(p,q) \over \d V_1(x(r))}\,dx(r) &=& \sum_\alpha
\Res_{\xi\rightarrow \mu_\alpha} {dE_{\xi,\overline{\xi}}(q)
B(p,\overline{\xi}) B(\xi,r) \over \left( y(\overline{\xi})-y(\xi)
\right) dx(\xi)}\cr
&=& \sum_\alpha \Res_{\xi\rightarrow
\mu_\alpha} {1 \over 2} {dE_{\xi,\overline{\xi}}(q)
\left[B(p,\overline{\xi}) B(\xi,r) + B(r,\overline{\xi}) B(\xi,p)
\right] \over \left( y(\overline{\xi})-y(\xi) \right) dx(\xi)}.\cr
\eea
Note that this corresponds exactly to the conjecture for
$k=1$ and $h=0$.

By integrating this expression with respect to $p$, we obtain the action on the
arrowed edges:
\beq\label{difE} {\d dE_{p,\overline{p}}(q) \over
\d V_1(x(r))}\,dx(r) = \sum_\alpha \Res_{\xi\rightarrow \mu_\alpha} {1
\over 2} {dE_{\xi,\overline{\xi}}(q)
\left[dE_{p,\overline{p}}(\overline{\xi}) B(\xi,r) +
B(r,\overline{\xi}) dE_{p,\overline{p}}(\xi) \right] \over \left(
y(\overline{\xi})-y(\xi) \right) dx(\xi)},
\eeq
where $q$, $p$ and $r$ are outside the integration contour.

There is one last quantity to evaluate, corresponding to the vertices:
\beq\label{difY}
{\d \over \d V_1(r)} \left( {1 \over y(p)-y(\overline{p})} \right)\,dx(r) = {B(p,r) - B(\overline{p},r) \over (y(p)-y(\overline{p}))^2 dx(p)}
\eeq

Let us now interpret diagrammatically these relations.

Eq. (\ref{difB}) represents the action of the loop insertion
operator on the non arrowed edges:
\beq
\begin{array}{l} {\epsfxsize 3cm\epsffile{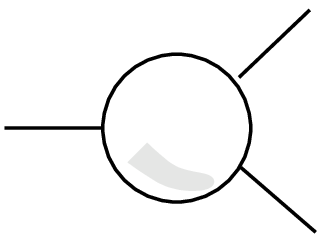}}\end{array}
={\d \over \d V_1} \begin{array}{l} {\epsfxsize 2.4cm\epsffile{berg.eps}}\end{array}
= \begin{array}{l} {\epsfxsize 2.5cm\epsffile{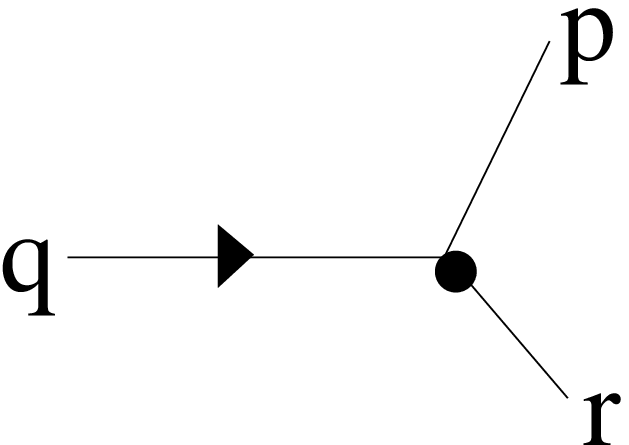}}\end{array}
+ \begin{array}{l} {\epsfxsize 2.5cm\epsffile{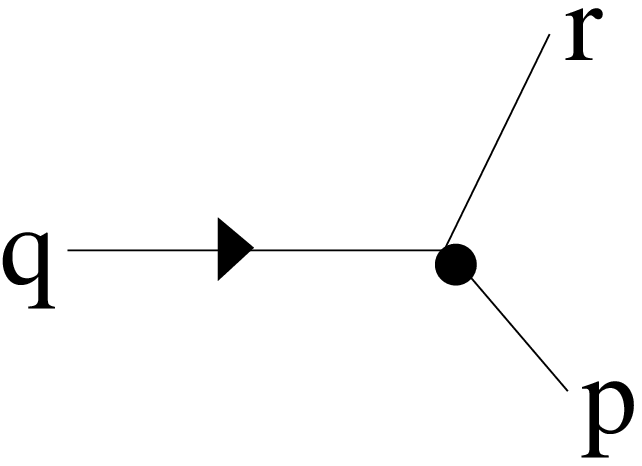}}\end{array}
.\eeq

In order to interpret \eq{difE}, one has to move the point $p$
inside the integration contour for $\xi$. For this purpose, let us
consider the action of the loop operator on the vertex, $dx(r)\,{\partial
\over \partial V_1(x(r))} \sum_\alpha \Res_{q\to \mu_\alpha}
{dE_{q,\overline{q}}(p) \over 2} {1 \over (y(\overline{q}) - y(q))
dx(q)}$. It gives the contribution
\beq\label{deriv} \begin{array}{l}  \sum_\alpha \Res_{q \to
\mu_\alpha} {dE_{q,\overline{q}}(p) \over 2} {B(\overline{q},r) -
B(q,r) \over (y(\overline{q}) - y(q))^2 dx(q)^2} \cr
 + \sum_\alpha \Res_{q \to \mu_\alpha} \Res_{\xi \to \mu_\alpha} {dE_{\xi,\overline{\xi}}(p) \over 2 (y(\overline{\xi})-y(\xi)) dx(\xi)}
 {dE_{q,\overline{q}}(\overline{\xi}) B(\xi,r) + dE_{q,\overline{q}}(\xi) B(\overline{\xi},r) \over 2 (y(\overline{q})- y(q))
 dx(q)}\cr\end{array}
\eeq
where $q$ lies outside the integration contour for $\xi$. We
move the integration contour for $\xi$ in the second term so that
$q$ lies inside.
One has poles only when $\xi \to q$ and $\xi \to
\overline{q}$ which can be written \beq \Res_{q \to \mu_\alpha}
\Res_{\xi \to \mu_\alpha} = \Res_{\xi \to \mu_\alpha} \Res_{q \to
\mu_\alpha} - \Res_{q \to \mu_\alpha} \Res_{\xi \to q} - \Res_{q
\to \mu_\alpha} \Res_{\xi \to \overline{q}}. \eeq
These contributions cancel totally the first term in \eq{deriv}.
Thus one finally obtains:
\bea &&dx(r)\,{\partial \over \partial V_1(x(r))} \sum_\alpha \Res_{q\to \mu_\alpha}
{dE_{q,\overline{q}}(p) \over 2} {1 \over (y(\overline{q}) - y(q))
dx(q)} = \cr && = \sum_\alpha \Res_{\xi \to \mu_\alpha} \Res_{q
\to \mu_\alpha} {dE_{\xi,\overline{\xi}}(p) \over 2
(y(\overline{\xi})-y(\xi)) dx(\xi)}
 {dE_{q,\overline{q}}(\overline{\xi}) B(\xi,r) + dE_{q,\overline{q}}(\xi) B(\overline{\xi},r) \over 2 (y(\overline{q})- y(q)) dx(q)},\cr
\eea

which corresponds to adding one leg to the vertex:
\beq
dx(r)\,{\partial \over \partial V_1(x(r))}\,
\begin{array}{l} {\epsfxsize 2.4cm\epsffile{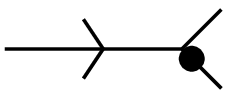}}\end{array}
=
\begin{array}{l} {\epsfxsize 8cm\epsffile{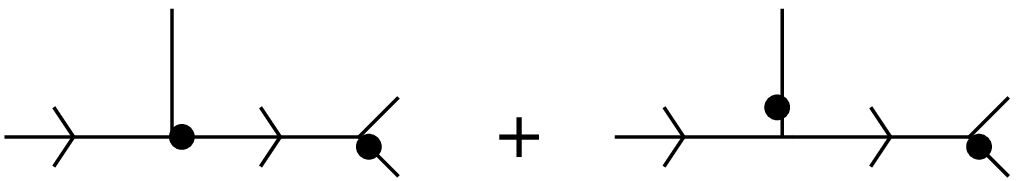}}\end{array}
\eeq
This ensures that the theorem is true.
}

Therefore we have the following diagrammatic rules:
\bt
The $k+1$ point function to order $h$, $W_{k+1}^{(h)}$ is the sum over all possible diagrams obtained as follows:
\begin{itemize}

\item choose one of the variables, say $p_1$ as the root.

\item draw all rooted skeleton binary trees (i.e. trees with vertices of valence $\leq 3$) with $k-1+2h$ edges. Draw arrows going from root to leaves, that puts a partial ordering on vertices.
A vertex $V_1$ preceeds $V_2$ if there is an oriented path from $V_1$ to $V_2$, and a vertex always preceeds itself.

\item add, in  all possible ways, $k$ non arrowed edges ending at the point $p_2,\dots, p_{k+1}$, so that each vertex has valence $\leq 3$.

\item complete the diagrams in all possible ways with $h$ non arrowed edges joining 2 comparable (with the partial ordering of  the tree) vertices, so as to get diagrams with valence $3$ only.

\item At each vertex, mark one leg with a dot, in all possible inequivalent ways. In general there are 2 possibilities at each vertex, except if the 2 branches coming out of the vertex are symetric (either 1 non-oriented edge, or 2 identical subgraphs with no external legs).
Each inequivalent diagram is counted exactly once (this is the standard way of computing symmetry factors).

\item assign to each such diagram a value obtained as follows: each non-arrowed edge is a Bergmann kernel, a residue of the form \eq{vertexrule} is computed at each vertex, where the $\ovl{q}$ variable corresponds to the marked edge.
The ordering for computing residues is given by the arrows, from leaves to root.

\end{itemize}

\et

Notice that those rules are extremely similar to those first found for the 1-matrix model in \cite{eynloop1mat, EyOran, ChEy}.
The 1-MM is merely the reduction of those rules to the case of an hyperelliptical surfaces.

\section{Topological expansion of the free energy}

In order to find the topological expansion of the free energy, one
has to "integrate wrt the potential $V_1$" the expansion of the
one point function.
First let us define the $H_x$ and $H_y$
operators corresponding respectively to the "inverse" of the loop
insertion operator ${\partial \over \partial V_1}$ and ${\partial
\over
\partial V_2}$.

\subsection{The $H$ operators}

We define the operators $H_x$ and $H_y$ such that, for any
meromorphic differential form $\phi$, one has
\beq H_x. \phi :=
\Res_{\infty_x} V_1(x)\,\phi -\Res_{\infty_y} (V_2(y)-xy)\,\phi +
T \int_{\infty_x}^{\infty_y} \phi + \sum_i \epsilon_i \oint_{{\cal
B}_i} \phi \eeq and \beq H_y. \phi := \Res_{\infty_y} V_2(y)\,\phi
-\Res_{\infty_x} (V_1(x)-xy)\,\phi + T \int_{\infty_y}^{\infty_x}
\phi -  \sum_i \epsilon_i \oint_{{\cal B}_i} \phi .\eeq

Note that \beq \label{property} (H_x+H_y).\phi =
\Res_{\infty_x,\infty_y} xy.\phi \eeq

Note that if $\phi=df$ is an exact differential, we have:
\beq
H_x. df = - \Res_{\infty_x,\infty_y} f\, y dx
\virg
H_y. df = - \Res_{\infty_x,\infty_y} f\, x dy.
\eeq

We now compute the action of $H_x$ on the two-point function on
the Bergmann kernel $B(p,q)$ (in this computation $H_x$ acts on
the variable $p$), that is
\bea H_x.B(p,q) &=& \Res_{\infty_x}
V_1(x(p))\,B(p,q) -\Res_{\infty_y} (V_2(y(p))-xy)\,B(p,q) \cr && +
T \int_{\infty_x}^{\infty_y} B(p,q) + \sum_i \epsilon_i
\oint_{{\cal B}_i} B(p,q).\cr\eea We integrate by parts the
residues at infinities. Provided that $B(p,q) = d_p dS_{p,o}(q)$,
that $V_1'(x(p))-y(p) \sim {T \over x(p)}$ as $p \to \infty_x$ and
that $V_2'(y(p))-x(p) \sim {T \over y(p)}$ as $p \to \infty_y$,
we can write \bea H_x.B(p,q) &=& -\Res_{\infty_x} {T\over
x(p)}dx(p)\,dS_{p,o}(q) -\Res_{\infty_x,\infty_y} y(p)
dx(p)\,dS_{p,o}(q)  \cr &&+\Res_{\infty_y} {T\over y(p)}dy(p)
\,dS_{p,o}(q) + T dE_{\infty_y,\infty_x}(q)  + \sum_i \epsilon_i
\oint_{{\cal B}_i} B(p,q)  .\cr \eea Finally, the Riemann bilinear
identities \cite{Farkas, Fay} associated to the fixed filling fractions
conditions give
\bea
H_X.B(p,q) &=& T dS_{\infty_x,o}(q)
-\Res_{\infty_x} y(p) dx(p)\,dS_{p,o}(q) - \Res_{\infty_y}
y(p)dx(p) \,dS_{p,o}(q) \cr && -T \,dS_{\infty_y,o}(q) + T
dS_{\infty_y,\infty_x}(q)  + \sum_i \epsilon_i \oint_{{\cal B}_i}
B(p,q)  \cr &=& \Res_{q} y(p) dx\,dS_{p,o}(q) \cr && - {1\over
2i\pi}\sum_i \oint_{{\cal A}_i} y(p)dx(p) \oint_{{\cal B}_i}
B(p,q)  + {1\over 2i\pi}\sum_i \oint_{{\cal B}_i} y(p) dx(p)
\oint_{{\cal A}_i} B(p,q) \cr &&   + \sum_i \epsilon_i
\oint_{{\cal B}_i} B(p,q)  \cr &=& -y(q)dx(q). \eea
That is:
\beq
H_x.B(.,q) = -y(q) dx(q) .
\eeq

\br The leading order of the free energy is already known~\cite{MarcoF}:
\bea 2{\cal F}^{(0)} &=& \Res_{\infty_x}
(V_1(x)+V_2(y)-xy)\,ydx + T \Pint_{\infty_x}^{\infty_y} ydx +
\sum_i \epsilon_i \oint_{{\cal B}_i} ydx \cr
 &=& \Res_{\infty_y} (V_1(x)+V_2(y)-xy)\,xdy + T \Pint_{\infty_y}^{\infty_x} xdy - \sum_i \epsilon_i \oint_{{\cal B}_i} xdy .\cr
\eea

In other terms, one has \beq 2{\cal F}^{(0)} = H_x(y dx) \eeq

\er

\subsection{Finding the free energy}

\bt For any $h$ we have :

 \beq\label{free1} (1-2h) W_{1,0}^{(h)} = H_x.
W_{2,0}^{(h)} \eeq \et

 Notice that this is already proven for $h=0$.

\proof{ Let us study how $H_x$ acts on the diagrams composing
$W_{2,0}^{(h)}$. By symmetry, one can consider that $H_x$ acts on
the leaf of the diagram\footnote{The diagrams composing
$W_{2,0}^{(h)}$ have two external legs, one root with an arrow
and one leaf which is a Bergmann kernel.}. Two
different configurations can occur.

If the leaf is linked to an innermost vertex, one has to compute
\bea \sum_{\alpha} \Res_{r \to \mu_\alpha} {dE_{r,\overline{r}}(q)
H_x.B(r,.) B(r,p) f(p) \over 2 ((y(\overline{r})-y(r)) dx(r)} &=&
- \sum_\alpha \Res_{r \to \mu_\alpha} {dE_{r,\overline{r}}(q) y(r)
B(r,p) f(p) \over 2 ((y(\overline{r})-y(r))} \cr &=& 0. \cr \eea

Otherwise, $H_x$ acts on $B(r,.)$ where there is an arrow going
out of the vertex corresponding to the integration of $r$:

\beq A = \sum_{\alpha'} \Res_{r \to \mu_{\alpha'}} \sum_{\alpha}
\Res_{p \to \mu_\alpha} {dE_{r, \overline{r}}(q)
dE_{p,\overline{p}}(r) y(\overline{r}) f(p) \over 2
(y(r)-y(\overline{r})) (y(p)-y(\overline{p})) dx(r) dx(p)} .\eeq

By moving the integration contours so that $r$ lies inside the
contour of $p$, one keeps only the contributions when $r\to p$ and
$r \to \overline{p}$, that is \bea A &=& \sum_{\alpha} \Res_{p \to
\mu_\alpha} \Res_{r \to p} {dE_{r, \overline{r}}(q)
dE_{p,\overline{p}}(r) y(\overline{r}) f(p) \over 2
(y(r)-y(\overline{r})) (y(p)-y(\overline{p})) dx(r) dx(p)}\cr && +
\sum_{\alpha} \Res_{p \to \mu_\alpha} \Res_{r \to \overline{p}}
{dE_{r, \overline{r}}(q) dE_{p,\overline{p}}(r) y(\overline{r})
f(p) \over 2 (y(r)-y(\overline{r})) (y(p)-y(\overline{p})) dx(r)
dx(p)} \cr &=& \sum_{\alpha} \Res_{p \to \mu_\alpha}
{dE_{p,\overline{p}}(q) f(p) \over 2 (y(p)-y(\overline{p})) dx(p)}
.\eea

The same diagram with $H_x$ acting on $B(\overline{r},.)$ gives
the same contribution and cancels the ${1 \over 2}$ factor.

Diagrammatically, this reads

\beq
\begin{array}{l}
{\epsfxsize 3cm\epsffile{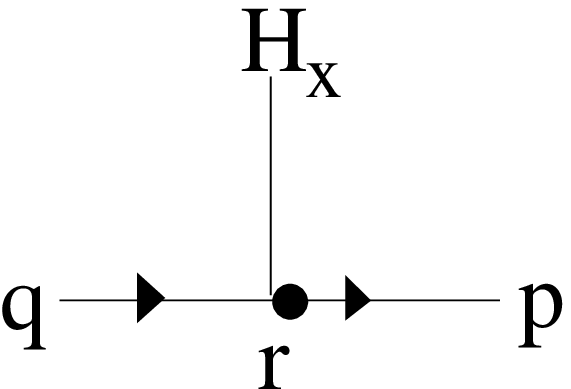}}
\end{array}
+ \begin{array}{l} {\epsfxsize 3cm\epsffile{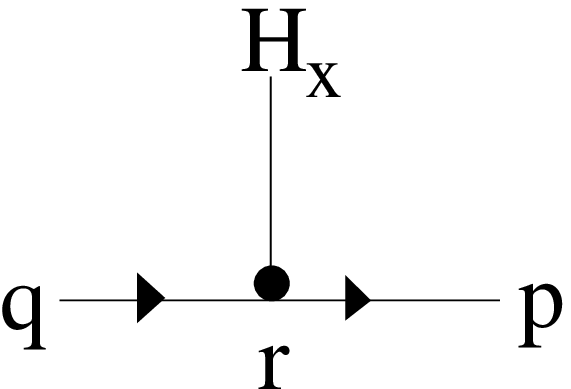}}
\end{array}= -
\begin{array}{l}
{\epsfxsize 3cm\epsffile{arrow.eps}}
\end{array}
\eeq

and

\beq
\begin{array}{l}
{\epsfxsize 3cm\epsffile{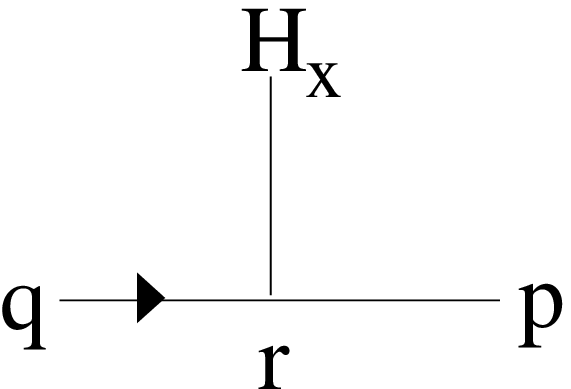}}
\end{array}
= 0. \eeq

Thus, because there are $2h-1$ arrowed edges composing any graph
contributing to $W_1^{(h)}$, one has: \bea
H_{x_q}.W_{2,0}^{(h)}(p,q) &=& H_{x_q}.{\partial \over \partial
V_1(q)} W_{1,0}^{(h)}(p) \cr &=& -(2h-1) W_{1,0}^{(h)}(p). \eea }

Beside, we have the following property:
\bl For any $h \neq 1$:
 \beq H_{x} . W_{1,0}^{(h)}(q) - H_{y} . W_{0,1}^{(h)}(q) = 0
 \eeq
\el
whose rather technical proof is found in appendix B.

\bc One easily derive from this theorem that for any $h\neq 1$,
the free energy can be written as
\beq
\encadremath{
(2-2h){\cal F}^{(h)}= - H_x W_{1,0}^{(h)}= - H_y W_{0,1}^{(h)},
}\eeq
up to an integration constant which does not depend on $V_1$ or $V_2$.
\ec

\subsection{Dependence on other parameters}

Now, with the explicit dependence of the free energy on the
potentials $V_1$ and $V_2$ in hands, one is interested in the
dependence in the other momenta, i.e. the filling fractions
$\epsilon_i$ and the temperature $T$.

The definitions of the filling fractions implies that \beq
{\partial y(p) dx(p) \over \partial \epsilon_i} = 2 i \pi du_i =
\oint_{{\cal{B}}_i} B(p,q), \eeq
where the $du_i$'s denote the
normalized holomorphic differential on cycles ${\cal A}_i$. This allows to
check that the derivation wrt the filling fractions is compatible
with our diagrammatic rules through \beq {\partial B(p,q) \over
\partial \epsilon_i} = \oint_{{\cal{B}}_i} {\partial \over
\partial V_1(x(p))} B(q,r) dx(q), \eeq which generalizes to \beq
\label{dereps} {\partial W_{1,0}^{(h)}(p) \over \partial
\epsilon_i} = \oint_{{\cal{B}}_i} W_{2,0}^{(h)}(p,q) .\eeq

In the same fashion, one derives \beq \label{derT} {\partial
W_{1,0}^{(h)}(p) \over \partial T} = \int_{\infty_x}^{\infty_y}
W_{2,0}^{(h)}(p,q) .\eeq

\subsection{Scaling equation}
For the sake of completeness and in order to interpret the formula
found for the free energy, one has to consider a slightly more
general model, in which the coupling constant between the two
matrices is not fixed, i.e. we consider the partition function
\beq Z:=\int_{H_N\times H_N} dM_1 dM_2\, e^{-{1\over \hbar}
\Tr(V_1(M_1) + V_2(M_2) - \kappa M_1 M_2 )} = e^{-{\cal F}}. \eeq
where $\kappa$ is an arbitrary non vanishing coupling constant.

One can easily obtain this model by rescaling our momenta: \beq
t_k \to {t_k \over \kappa} \;\;\hbox{,}\;\; \tilde{t}_k \to
{\tilde{t}_k \over \kappa} \;\;\hbox{and} \;\; T \to {T \over
\kappa}. \eeq

After this redefinition of the parameters, all the results
presented in this paper remain valid\footnote{ Note that the
diagrammatic rules remain the same, only the Riemann surface is
affected by this rescaling. } and one obtains \beq (2-2h) {\cal
F}^{(h)} = -H_x W_{1,0}^{(h)} + f(T,\epsilon_i,\kappa),\eeq with
\beq H_x. \phi := \Res_{\infty_x} V_1(x) \,\phi -\Res_{\infty_y}
(V_2(y)- \kappa xy)\,\phi + T \int_{\infty_x}^{\infty_y} \phi +
\sum_i \epsilon_i \oint_{{\cal B}_i} \phi .\eeq

Now, define \beq \forall h \neq 1 \;\; \tilde{F}^{(h)} = - {H_x
W_{1,0}^{(h)}(p) \over 2-2h} .\eeq

Eq. (\ref{dereps}) and \eq{derT} give

\bea (2-2h) {\partial \tilde{F}^{(h)} \over \partial \epsilon_i}
&=& - {\partial H_x W_{1,0}^{(h)}(p) \over \partial \epsilon_i}
\cr &=& - \left({\partial H_x \over \partial \epsilon_i}\right)
W_{1,0}^{(h)}(p) - H_x {\partial W_{1,0}^{(h)}(p) \over \partial
\epsilon_i} \cr &=& - (2-2h) \oint_{{\cal{B}}_i}
W_{1,0}^{(h)}(p).\cr \eea

That is to say \beq {\partial \tilde{F}^{(h)} \over \partial
\epsilon_i} = - \oint_{{\cal{B}}_i} W_{1,0}^{(h)}(p) \;\;\;
\hbox{and} \;\;\; {\partial \tilde{F}^{(h)} \over \partial T} = -
\int_{\infty_x}^{\infty_y} W_{1,0}^{(h)}(p) .\eeq

On the other hand, by definition \beq \sum_{k=1}^{d_1} t_k {\d \tilde{F}^{(h)} \over \d t_k} = \Res_{\infty_x} V_1(x(p)) {
\d \tilde{F}^{(h)} \over \d V_1(x(p))}.\eeq Which leads to \beq
\sum_{k=1}^{d_1} t_k {\d \tilde{F}^{(h)} \over \d t_k} =
- \Res_{\infty_x} V_1(x(p)) W_{1,0}^{(h)}(p) .\eeq

One can also derive \bea \sum_{k=1}^{d_2} \tilde{t}_k
{\d \tilde{F}^{(h)} \over \d \tilde{t}_k} &=& - \Res_{\infty_y}
{V_2(y(p)) \over (2-2h)} {\d H_x.W_{1,0}^{(h)}(.) \over \d
V_2(y(p))}\cr &=&  - \Res_{\infty_y} {V_2(y(p)) \over (2-2h)} {\d
(H_x+H_y).W_{1,0}^{(h)}(.) \over \d V_2(y(p))} \cr && +
\Res_{\infty_y} {V_2(y(p)) \over (2-2h)} {\d H_y.W_{1,0}^{(h)}(.)
\over \d V_2(y(p))} \cr &=& \Res_{\infty_y} V_2(y(p))
W_{1,0}^{(h)}(p).\cr\eea The last identity uses the property
\eq{property} of the $H$ operators.

There is one additive momentum that one has to consider now, which
gives \beq {\d \tilde{F}^{(h)} \over \d \kappa} = -
\Res_{\infty_y} x(p) y(p) W_{0,1}(p)\eeq

Thus, putting everything together, for any set of filling
fractions $\epsilon_i$, any coupling constant $\kappa$ and any
temperature $T$, $\tilde{F}^{(h)}$ fulfills the homogenous
equation
\beq \label{homogenous}
\encadremath{
(2-2h)\tilde{F}^{(h)} =
\sum_{k=1}^{d_1+1} t_k  {\partial \tilde{F}^{(h)} \over
\partial t_k} + \sum_{k=1}^{d_2+1} \tilde{t}_k {\partial
\tilde{F}^{(h)} \over
\partial \tilde{t}_k} + T {\partial \tilde{F}^{(h)} \over \partial T} + \kappa {\partial \tilde{F}^{(h)} \over \partial \kappa} + \sum_i \epsilon_i {\partial
\tilde{F}^{(h)} \over \partial \epsilon_i} .
}\eeq

Hence, $\tilde{F}^{(h)}$ is our final answer for the free energy:

\beq \label{final} \tilde{F}^{(h)}={\cal F}^{(h)} = {1 \over 2h-2}
H_x.W_{(1,0)}^{(h)} = {1 \over 2h-2} H_y.W_{(0,1)}^{(h)} .
\eeq

\br One can note that \eq{homogenous} is nothing but saying that
rescaling the expansion parameter $\hbar$ is equivalent to
rescaling all the other momenta in the same way:
\beq
0= \hbar {\partial \tilde{F} \over \partial \hbar}
 + \sum_{k=1}^{d_1+1} t_k {\partial \tilde{F} \over \partial t_k}
 + \sum_{k=1}^{d_2+1} \tilde{t}_k {\partial \tilde{F} \over \partial \tilde{t}_k}
 + T {\partial \tilde{F} \over \partial T}
 + \kappa {\partial \tilde{F} \over \partial \kappa}
 + \sum_i \epsilon_i {\partial \tilde{F} \over \partial \epsilon_i} .
 \eeq
 \er

\subsection{Explicit computation of the free energy: diagrammatic
rules}

The equation (\ref{final}) allows to compute any term in the
expansion of the free energy, but the first correction to the
leading order, using diagrammatic rules. For this purpose, one
just introduce a new bi-valent vertex corresponding to the action
of the $H_x$ operator on the root of the diagrams composing the
one point functions. One represent it, \beq \begin{array}{l}
{\epsfxsize 2.3cm\epsffile{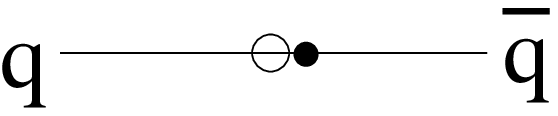}}\end{array} = \sum_\alpha
\Res_{q\to \mu_\alpha} {- {1\over 2} \int_{\overline{q}}^q y(\xi)
dx(\xi) \over (y(q)-y(\overline{q}))dx(q)} \eeq Using this new
vertex, one obtains the $h$'th order correction to the free energy
by considering the sum of the diagrams contributing to
$W_{1,0}^{(h)}$ where the ingoing arrow is replaced by this new
bi-valent vertex.

\br We must remind that this technic does not give the first order
correction. Nevertheless this particular correction has already
been computed in the literature~\cite{EKK}.
 \er

\section{Conclusion}

In this paper, we have obtained a closed expression for the complete expansion of the free energy of the formal hermitian two-matrix
model as residues on the spectral curve giving an answer to one of the problems left undone in \cite{EyOran}.
On the way, we also refined the diagrammatic rules used to compute non-mixed correlation functions. We particularly
exhibit that these functions actually only depend on what happens near the branch points. The link with the
diagrammatic technique of \cite{eynloop1mat} for the 1MM looks also more evident.

Nevertheless, there still remain many different problems one should address from this starting point. In this paper,
we begin the computation of correlation functions mixing traces in $M_1$ and traces in $M_2$, but the general case
does not seem as simple for the moment. In the same way, using the result of \cite{EOtrmixte}, one would like to
compute the expansion of mixed traces.

This natural generalization of diagrammatic techniques from the 1MM to the 2MM also points out that the study
of the chain of matrices \cite{eynmultimat} would give a more general knowledge of the link between algebraic geometry and matrix models.

Let us also mention that we have considered here only polynomial potentials, but it is clear that the whole method should extend easily to
the more general class of semi-classical potentials \cite{marcosemipot, Bertolasem}, whose loop equations are very similar \cite{eylooprat}.
The 1-MM with hard edges was already treated in \cite{chhard}.

Another extension of this problem is to compute the $1/N$ expansion of random matrix models for non-hermitian matrices. This problem is already partialy addressed in \cite{WZ1,WZ2,WZ3},
but a diagrammatic technique has not been found yet.

Finally, as we already mentioned it, solving the formal model is different from solving the physical problem
corresponding to a convergent integral and constrained filling fractions lowering the energy.
This problem, which is expected to be similar to \cite{BDE} is not addressed for the moment and should give a natural complement to this work.

\subsection*{Aknowledgements}
We are grateful to A. ZAbrodin for discussions.
L. Chekhov thanks RFBR grant No 05-01-00498, grant for Support Scientific Schools 2052.2003.1, the program Mathematical Methods of Nonlinear Dynamics.
All the authors thank the Enigma european network MRT-CT-2004-5652, and the ANR project G\'eom\'etrie et int\'egrabilit\'e en physique math\'ematique ANR-05-BLAN-0029-01.


\setcounter{section}{0}

\appendix{Computation of $ W_{k,1}^{(h)}$.}

In this section, we compute a new set of correlation functions
including one trace in $M_2$. In other words, we compute all the $
W_{k,1}^{(h)}$'s. More precisely, we explain how the diagrammatic
rules are modified when one changes one trace \beq \Tr {1 \over
x(p) - M_1} \; \to \; \Tr {1 \over y(p) - M_2} .\eeq

We do not give explicit formulas for all the  $W_{k,l}^{(h)}$ with
$l > 1$. This problem will be addressed in another work.

\subsection{Examples}

In order to give some idea of the result, let us review some
already known functions~\cite{eynm2mg1,Bertola,Kri}.

\subsubsection{2-point functions}

\beq W_{2,0}^{(0)}(p,q) =W_{0,2}^{(0)}(p,q) =-W_{1,1}^{(0)}(p,q)
=B(p,q) \eeq

\subsubsection{3-point functions}

\beq W_{3,0}^{(0)}(p_1,p_2,p_3) = \sum_{alpha} \Res_{q\to
\mu_\alpha} {B(p_1,q)B(p_2,q)B(p_3,q)\over dx(q)dy(q)} \eeq \beq
\label{w21} W_{2,1}^{(0)}(p_1,p_2,p_3) = - \sum_{\alpha} \Res_{q\to
\mu_\alpha} {B(p_1,q)B(p_2,q)B(p_3,q)\over dx(q)dy(q)}
 - \Res_{q\to p_3} {B(p_1,q)B(p_2,q)B(p_3,q)\over dx(q)dy(q)}
\eeq

\proof{ \bea W_{2,1}(p_1,p_2,q) &=& {\partial \over \partial
V_1(x(p_2))}(w_{1,1}(p_1,q)) dy(q) dx(p_1) dx(p_2) \cr &=& -
{\partial \over \partial V_1(x(p_2))}(w_{2,0}(p_1,q) {dx(q) \over
dy(q)}) dy(q) dx(p_1) dx(p_2) \cr &=& - W_{3,0}(p_1,p_2,q) +
{W_{2,0}(p_1,q) \over dx(q)} d_q\left( {W_{1,1}(p_2,q) \over
dy(q)} \right) \cr && + {W_{1,1}(p_2,q) \over dy(q)} d_q \left(
{W_{2,0}(p_1,q) \over dx(q)} \right) \cr &=&  - W_{3,0}(p_1,p_2,q)
+d_q \left( {W_{2,0}(p_1,q) W_{1,1}(p_2,q) \over dx(q) dy(q)}
\right) \cr \eea

}

\subsection{General case }

\bt For any $|K|+h >1$

\beq W_{|K|,1}^{(h)}(p_1,\dots,p_k,q) +
W_{|K|+1,0}^{(h)}(p_1,\dots,p_k,q) = d_q f^{(h)}(p_1,\dots,p_k,q)
\eeq where $f^{(h)}(p_1,\dots,p_k,q)$ is obtained from the
diagrams composing $W_{|K|,0}^{(h)}(p_1,\dots,p_k)$ by cutting any
of its propagator as follows:

\beq  \begin{array}{l} {\epsfxsize 2.4cm\epsffile{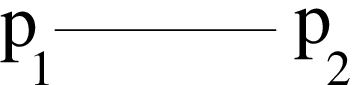}}
\end{array}
\to {1 \over dx(q) dy(q)} \begin{array}{l} {\epsfxsize
5cm\epsffile{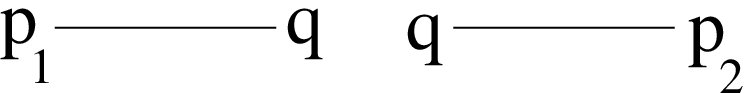}}
\end{array}
\eeq

and

\beq  \begin{array}{l} {\epsfxsize 2.4cm\epsffile{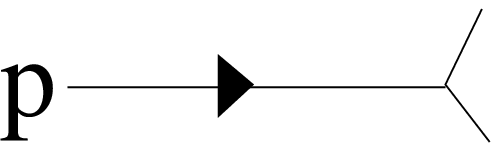}}
\end{array}
\to {1 \over dx(q) dy(q)} \begin{array}{l} {\epsfxsize
5cm\epsffile{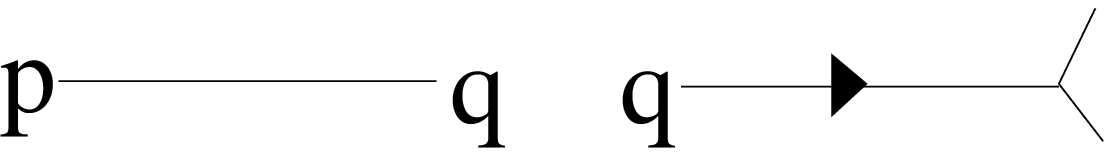}}
\end{array}
.\eeq

\et

\proof{

$W_{|K|,1}^{(h)}(p_K,q)$ can be obtained by differentiating
$W_{|K|,0}^{(h)}(p_K)$ wrt $V_2(y(q))$. So one has to know the
action of ${\partial \over \partial V_2}$ on the diagrammatic
rules.

One can reexpress the 3-point function of \eq{w21} as \beq
{\partial \over
\partial V_2(y(s))} B(q,p)
= - {\partial \over \partial V_1(x(s))} B(q,p) - d_s \left(
{B(s,p) B(s,q) \over dx(s) dy(s)} \right) .\eeq

This can be graphically represented by

\bea {\d \over \d V_2(y(s))} \begin{array}{l} {\epsfxsize
2.3cm\epsffile{berg.eps}}
\end{array}
&=& - {\d \over \d V_1(x(s))} \begin{array}{l} {\epsfxsize
2.4cm\epsffile{berg.eps}}
\end{array}\cr
&& - d_s\left({1 \over dx(s) dy(s)} \begin{array}{l} {\epsfxsize
4.9cm\epsffile{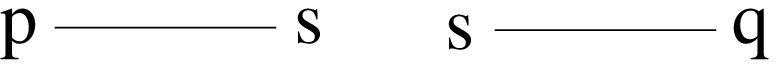}}
\end{array}\right) . \cr
\eea

On the other hand, by integration, this implies that \bea
{\partial \over \partial V_2(y(s))} dE_{p,r}(q) &=& - \sum_\alpha
\Res_{a \to \mu_\alpha} {dE_{a,\overline{a}}(q) \left[
dE_{p,r}(\overline{a}) B(a,s) + B(s,\overline{a}) dE_{p,r}(a)
\right] \over 2 (y(\overline{a})-y(a)) dx(a)}\cr && - \sum_\alpha
\Res_{a \to \mu_\alpha} \Res_{b \to a} {dE_{a,b}(q) dE_{p,r}(b)
B(a,s) \over (y(b)-y(a)) (x(b)-x(a))}.\cr \eea

Thus the derivative wrt $V_2$ of the first vertex, ${\partial
\over \partial V_2(y(s))} \sum_\alpha \Res_{p\to \mu_\alpha}
{dE_{p,\overline{p}}(q) \over 2 (y(\overline{p}) - y(p)) dx(p)}$
gives: \beq \begin{array}{l}  - \sum_\alpha \sum_\beta \Res_{p\to
\mu_\alpha} \Res_{a \to \mu_\beta} {dE_{a,\overline{a}}(q) \left[
dE_{p,\overline{p}}(\overline{a}) B(a,s) + B(s,\overline{a})
dE_{p,\overline{p}}(a) \right] \over 4 (y(\overline{a})-y(a))
dx(a) (y(\overline{p}) - y(p)) dx(p)}\cr  - \sum_\alpha \Res_{p\to
\mu_\alpha} \Res_{a \to s} \Res_{b \to a} {dE_{a,b}(q)
dE_{p,\overline{p}}(b) B(a,s) \over 2 (y(b)-y(a)) (x(b)-x(a))
(y(\overline{p}) - y(p)) dx(p) }\cr + \sum_\alpha \Res_{p\to
\mu_\alpha} {dE_{p,\overline{p}}(q) \over 2} {\partial \over
\partial V_2(s)} {1 \over (y(\overline{p}) - y(p)) dx(p)} \cr
\end{array},
\eeq where $p$ lies outside the integration contour for $a$. Note
that all the derivations have been performed by keeping the $x$'s
fixed.

Moving the integration contours so that $p$ lies inside gives a
contribution that cancels the last terms and one obtains \beq
\begin{array}{rcl}
{\partial \over \partial V_2(y(s))} \sum_\alpha \Res_{p\to
\mu_\alpha} {dE_{p,\overline{p}}(q) \over 2 (y(\overline{p}) -
y(p)) dx(p)}&=& - {\partial \over
\partial V_1(x(s))} \sum_\alpha \Res_{p\to
\mu_\alpha} {dE_{p,\overline{p}}(q) \over 2 (y(\overline{p}) -
y(p)) dx(p)} \cr && - d_s \left( {B(s,q) \over dx(s) dy(s)}
\sum_\alpha \Res_{p\to \mu_\alpha} { dE_{p,\overline{p}}(s) \over
2 (y(\overline{p}) - y(p)) dx(p) }\right)\cr
\end{array}. \eeq

This graphically reads:

\bea {\d \over \d V_2(y(s))} \begin{array}{l} {\epsfxsize
2.4cm\epsffile{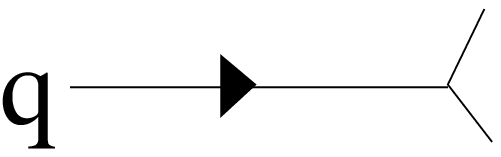}}
\end{array}
&=& - {\d \over \d V_1(x(s))} \begin{array}{l} {\epsfxsize
2.4cm\epsffile{vert1.eps}}
\end{array}\cr
&& - d_s\left({1 \over dx(s) dy(s)} \begin{array}{l} {\epsfxsize
5cm\epsffile{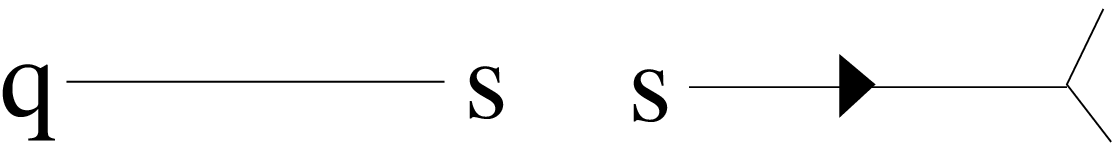}}
\end{array}\right) \cr
\eea

These two relations prove straightforwardly the theorem. }

\br For $|K|=1$, one can go further in the description of
$W_{1,1}^{(h)}(p_1,p_2) + W_{2,0}^{(h)}(p_1,p_2)$. Indeed, the
graphs obtained by cutting the propagators of $W_{1,0}^{(h)}(p_1)$
have no pole when $p_1 \to p_2$ except when one cuts the root of
the graph, in which case it gives the contribution ${B(p_1,p_2)
\over dx(p_2) dy(p_2)} W_{1,0}^{(h)}(p_2)$, that is to say:

\beq\label{remarkddv2k1} W_{1,1}^{(h)}(p_1,p_2) + W_{2,0}^{(h)}(p_1,p_2) = - d_{p_2}
\left\{ {B(p_1,p_2) \over dx(p_2) dy(p_2)} W_{1,0}^{(h)}(p_2) +
f(p_1,p_2) \right\}, \eeq where $f(p_1,p_2)$ has no pole as $p_1
\to p_2$.
 \er

\appendix{Symmetry of the free energy under the exchange of $x$ and $y$}

For $h \neq 1$

\beq H_{x_q} . W_{1,0}^{(h)}(q) - H_{y_q} . W_{0,1}^{(h)}(q) =
\left(H_{x_{p_1}} H_{y_{p_2}} - H_{y_{p_2}} H_{x_{p_1}} \right)
W_{1,1}^{(h)}(p_1,p_2) \eeq

The only terms that contributes in the exchange of $H_{x}$ and
$H_y$ come from the integration contours around $\infty_x$ and
$\infty_y$, i.e.  \bea\label{commut1} {\cal A}&=&\left(H_{x_{p_1}}
H_{y_{p_2}} - H_{y_{p_2}} H_{x_{p_1}} \right)
W_{1,1}^{(h)}(p_1,p_2) \cr &=& - \Res_{p_1 \to \infty_x} \Res_{p_2
\to p_1} V_1(x(p_1)) (V_1(x(p_2)) - x(p_2) y(p_2))
W_{1,1}^{(h)}(p_1,p_2)  - \cr && - \Res_{p_1 \to \infty_y}
\Res_{p_2 \to p_1} V_2(y(p_2)) (V_2(y(p_1)) - x(p_1) y(p_1))
W_{1,1}^{(h)}(p_1,p_2) .\cr \eea

On the other hand, one knows from remark \eq{remarkddv2k1}, that \beq W_{1,1}^{(h)}(p_1,p_2) = -
W_{2,0}^{(h)}(p_1,p_2) - d_{p_2} \left\{ {B(p_1,p_2) \over dx(p_2)
dy(p_2)} W_{1,0}^{(h)}(p_2) + f(p_1,p_2) \right\} .\eeq

where $f(p_1,p_2)$ has no pole when $p_2 \to p_1$. Note that
$W_{2,0}^{(h)}(p_1,p_2)$ do not have pole when $p_2 \to p_1$
either.

Thus, the only non vanishing terms in \eq{commut1} come from $
d_{p_2} \left\{ {B(p_1,p_2) \over dx(p_2) dy(p_2)}
W_{1,0}^{(h)}(p_2) \right\} $.

This reads \beq \begin{array}{l} \Res_{p_1 \to \infty_x} \Res_{p_2
\to p_1} V_1(x(p_1)) (V_1(x(p_2)) - x(p_2) y(p_2)) d_{p_2} \left\{
{B(p_1,p_2) \over dx(p_2) dy(p_2)} W_{1,0}^{(h)}(p_2) \right\} \cr
 + \Res_{p_1 \to \infty_y} \Res_{p_2 \to p_1} V_2(y(p_2))
(V_2(y(p_1)) - x(p_1) y(p_1)) d_{p_2} \left\{ {B(p_1,p_2) \over
dx(p_2) dy(p_2)} W_{1,0}^{(h)}(p_2) \right\} .\cr \end{array} \eeq

We now evaluate these residues by part and obtain that \bea
{\cal{A}} &=& - \Res_{p \to \infty_x} V_1(x(p)) d_p \left[
{(V_1'(x(p))-y(p)) dx(p) - x(p) dy(p) \over dx(p) dy(p) }
W_{1,0}^{(h)}(p) \right] \cr && - \Res_{p \to \infty_y} (V_2(y(p))
- x(p) y(p)) d_p \left[ {V_2'(y(p)) W_{1,0}^{(h)}(p) \over dx(p)}
\right] .\cr \eea

Another integration by part can be written \bea {\cal A} &=& +
\Res_{p \to \infty_x} V_1'(x(p)) {(V_1'(x(p))-y(p)) dx(p) - x(p)
dy(p) \over dy(p) } W_{1,0}^{(h)}(p) \cr && + \Res_{p \to
\infty_y} (V_2'(y(p)) dy(p) - x(p) dy(p) - y(p) dx(p)) {V_2'(y(p))
W_{1,0}^{(h)}(p) \over dx(p)} .\cr \eea

Knowing that $V_1'(x(p)) - y(p) \sim {T \over x(p)}$ when $ p \to
\infty_x $ and $V_2'(y(p)) - x(p) \sim {T\over y(p)} $ when $p \to
\infty_y$ and that the other factors do not have poles at
infinities, one can finally write that \beq \left(H_{x_{p_1}}
H_{y_{p_2}} - H_{y_{p_2}} H_{x_{p_1}} \right)
W_{1,1}^{(h)}(p_1,p_2) = - \Res_{q \to \infty_x \, , \,
\infty_{y}} x(q) y(q) W_{1,0}^{(h)}(q) .\eeq

Let us now show that the RHS vanishes by looking precisely at the
form of $W_{1,0}^{(h)}(q)$. Because it comes from a vertex,
\eq{conjecturerecW}, with moving the integration contour, implies
that \bea 2 \Res_{\infty_x \, , \, \infty_{y}} x(q) y(q)
W_{1,0}^{(h)}(q) &=& -2 \sum_\alpha \Res_{q \to \mu_\alpha} x(q)
y(q) W_{1,0}^{(h)}(q) \cr &=& - \sum_\alpha \sum_{\beta} \Res_{q
\to \mu_\alpha} \Res_{p \to \mu_\beta} {x(q) y(q)
dE_{p,\overline{p}}(q) \over (y(p)-y(\overline{p})) dx(p)} \times
 \cr && \times \left[ W_{2,0}^{(h-1)}(p,p) +
\sum_{m=1}^{h-1} W_{1,0}^{(m)}(p) W_{1,0}^{(h-m)}(p)\right] .\cr
\eea Now we move one more time the integration contour and clearly
segregate the case when $p$ and $q$ are near the same branch
point. \bea 2 \Res_{\infty_x \, , \, \infty_{y}} x(q) y(q)
W_{1,0}^{(h)}(q) &=& - \sum_{\alpha \neq beta} \Res_{p \to
\mu_\alpha} \Res_{q \to \mu_\beta} {x(q) y(q)
dE_{p,\overline{p}}(q) \over (y(p)-y(\overline{p})) dx(p)}
 \cr && \times \left[ W_{2,0}^{(h-1)}(p,p) +
\sum_{m=1}^{h-1} W_{1,0}^{(m)}(p) W_{1,0}^{(h-m)}(p)\right] - \cr
&& - \sum_\alpha \left[ \Res_{p \to \mu_\alpha} \Res_{q \to
\mu_\alpha} - \Res_{q \to \mu_\alpha} \Res_{p \to q \, , \,
\overline{q}}\right] {x(q) y(q) dE_{p,\overline{p}}(q) \over
(y(p)-y(\overline{p})) dx(p)} \cr && \times \left[
W_{2,0}^{(h-1)}(p,p) + \sum_{m=1}^{h-1} W_{1,0}^{(m)}(p)
W_{1,0}^{(h-m)}(p)\right] .\cr \eea

One can see that the integrand has no pole as $q$ approaches a
branch point. Thus one keeps only \beq \sum_\alpha \Res_{q \to
\mu_\alpha} \Res_{p \to q \, , \, \overline{q}} {x(q) y(q)
dE_{p,\overline{p}}(q) \over (y(p)-y(\overline{p})) dx(p)} \left[
W_{2,0}^{(h-1)}(p,p) + \sum_{m=1}^{h-1} W_{1,0}^{(m)}(p)
W_{1,0}^{(h-m)}(p)\right] .\eeq

We finally perform the integration and get \bea 2 \Res_{\infty_x
\, , \, \infty_{y}} x(q) y(q) W_{1,0}^{(h)}(q) &=& \sum_\alpha
\Res_{q \to \mu_\alpha} {x(q) \over dx(q)} \left[
W_{2,0}^{(h-1)}(p,p) + \sum_{m=1}^{h-1} W_{1,0}^{(m)}(p)
W_{1,0}^{(h-m)}(p)\right] \cr &=& 0 .\eea {\bf QED.}


\end{document}